\documentclass[5p,times]{elsarticle}

\usepackage{amssymb}
\usepackage{amsmath}

\usepackage{graphicx}
\usepackage[countmax]{subfloat}
\usepackage{float}
\usepackage{hyperref}
\usepackage{subfig}
\usepackage[ruled]{algorithm2e}
\usepackage{booktabs}
\usepackage{multirow}
\usepackage{caption}
\usepackage{wrapfig}
\usepackage{flushend}
\begin{document}

\begin{frontmatter}



\title{MLQM: Machine Learning Approach for Accelerating Optimal Qubit Mapping}


\author[1]{Wenjie Sun}
\ead{202211022528@std.uestc.edu.cn}

\author[2]{Xiaoyu Li\corref{cor1}}
\ead{xiaoyuuestc@uestc.edu.cn}

\author[3]{Lianhui Yu}
\ead{202222120303@std.uestc.edu.cn}

\author[1]{Zhigang Wang}
\ead{zhigangwang@uestc.edu.cn}

\author[4]{Geng Chen}
\ead{202312081614@uestc.edu.cn}

\author[4]{Guowu Yang}
\ead{guowu@uestc.edu.cn}

\cortext[cor1]{Corresponding author}

\affiliation[1]{organization={The School of Electronic Science and Engineering, University of Electronic Science and Technology of China},
            addressline={No.2006, Xiyuan Ave}, 
            city={Chengdu},
            postcode={611731}, 
            country={China}}
            
\affiliation[2]{organization={The School of Information and Software Engineering, University of Electronic Science and Technology of China},
	city={Chengdu},
	postcode={611731}, 
	country={China}}

\affiliation[3]{organization={The School of Physics, University of Electronic Science and Technology of China},
	city={Chengdu},
	postcode={611731}, 
	country={China}}

\affiliation[4]{organization={The School of Computer Science and Engineering, University of Electronic Science and Technology of China},
	city={Chengdu},
	postcode={611731}, 
	country={China}}

\begin{abstract}
Quantum circuit mapping is a critical process in quantum computing that involves adapting logical quantum circuits to adhere to hardware constraints, thereby generating physically executable quantum circuits. Current quantum circuit mapping techniques, such as solver-based methods, often encounter challenges related to slow solving speeds due to factors like redundant search iterations. Regarding this issue, we propose a machine learning approach for accelerating optimal qubit mapping (MLQM). First, the method proposes a global search space pruning scheme based on prior knowledge and machine learning, which in turn improves the solution efficiency. Second, to address the limited availability of effective samples in the learning task, MLQM introduces a novel data augmentation and refinement scheme, this scheme enhances the size and diversity of the quantum circuit dataset by exploiting gate allocation and qubit rearrangement. Finally, MLQM also further improves the solution efficiency by pruning the local search space, which is achieved through an adaptive dynamic adjustment mechanism of the solver variables. Compared to state-of-the-art qubit mapping approaches, MLQM achieves optimal qubit mapping with an average solving speed-up ratio of 1.79 and demonstrates an average advantage of 22\% in terms of space complexity.
\end{abstract}


\begin{highlights}
\item MLQM introduces a machine learning approach with novel data augmentation, significantly reducing the quantum circuit mapping search space.
\item An adaptive hyperparameter mechanism dynamically transforms search trends into constraints, enhancing mapping efficiency.
\item Experiments show MLQM achieves 1.79× solving time speedup and 22\% memory reduction, advancing quantum circuit optimization.

\end{highlights}


\end{frontmatter}



\begin{abstract}
	Quantum circuit mapping is a critical process in quantum computing that involves adapting logical quantum circuits to adhere to hardware constraints, thereby generating physically executable quantum circuits. Current quantum circuit mapping techniques, such as solver-based methods, often encounter challenges related to slow solving speeds due to factors like redundant search iterations. Regarding this issue, we propose a machine learning approach for accelerating optimal qubit mapping (MLQM). First, the method proposes a global search space pruning scheme based on prior knowledge and machine learning, which in turn improves the solution efficiency. Second, to mitigate the scarcity of effective samples in the learning task above, MLQM incorporates a data augmentation and refinement scheme. This scheme enhances the size and diversity of the quantum circuit dataset by exploiting gate assignment and qubit rearrangement. Finally, MLQM also further improves the solution efficiency by pruning the local search space, which is achieved through an adaptive dynamic adjustment mechanism of the solver variables. Compared to state-of-the-art qubit mapping approaches, MLQM achieves optimal qubit mapping with an average solving speed-up ratio of 1.79 and demonstrates an average advantage of 22\% in terms of space complexity.
\end{abstract}

\section{Introduction}
Quantum computing has the potential to solve complex problems that are intractable for classical computers \cite{Harrigan2021,Arute2020,McArdle2019,McArdle2019,Huang2024}, and its technical research and industrialization process have achieved initial successes in the NISQ era \cite{Jurcevic2021,Arute2019,Reagor2018}. Among the various technological approaches to implementing quantum computing \cite{Leon2021,Henriet2020,Arute2019}, superconducting quantum circuits have emerged as one of the most prevalent due to their advantages in designability and scalability \cite{Huang2020}. To ensure the correct execution of quantum programs on superconducting quantum computers, logic synthesis and qubit mapping are two fundamental processes. Logic synthesis involves decomposing high-level gates in quantum programs into single-qubit or two-qubit gates supported by the quantum hardware. Qubit mapping, also known as layout synthesis, is employed to insert swap gates into quantum circuits to satisfy the hardware connectivity constraints of the quantum computer. The evaluation metrics of qubit mapping include circuit depth, number of swap gates, fidelity, etc., and the optimal qubit mapping refers to a mapping in which one or more metrics above are optimal.

In previous research, methods for addressing the qubit mapping problem can be broadly categorized into two classes: heuristic methods and solver-based methods. Heuristic methods \cite{Lao2022,liTacklingQubitMapping2019,Zhou2020,Li2023a,Siraichi2019,Zulehner2019,liQubitMappingBased2021} utilize meta-heuristic algorithms to determine quantum circuit mapping schemes, with a notable example being the SABRE method proposed by Gushu Li et al. \cite{liTacklingQubitMapping2019}. While heuristic methods generally exhibit excellent execution speeds, the solutions obtained are often far from optimal and lack sufficient interpretability \cite{Cong2023}, leading to limited practicality.

Solver-based approaches to the qubit mapping problem involve transforming the problem into a satisfiability modulo theories (SMT) problem, encoding the objective as constraints, and employing solvers to search for solutions that satisfy the constraints \cite{Wu2022,Tan2021,Tan2020,Wille2014}. Robert Wille first proposed utilizing solver-based Pseudo Boolean Optimization to achieve the optimal number of swap gates in 2014 \cite{Wille2014}. In 2020, Bochen Tan et al. treated layout synthesis as a mathematical optimization problem and employed time-space variable encoding to obtain solutions with either optimal depth or optimal swap gate count \cite{Tan2020}. They also proposed a coarse-grained time model called TB-OLSQ, which reduced time consumption, but the optimality of the circuit depth could not be guaranteed. One year later, Bochen Tan et al. introduced a solver-based approach based on gate absorption methods, using solver constraints to find swap gates absorbable by $\text{SU}(4)$ gates \cite{Tan2021}. Qubit mapping approaches combined solver methods and heuristics were also proposed to address the problem. Abtin Molavi et al. proposed a qubit mapping method considering the problem as a MaxSAT problem \cite{Molavi2022}, and local relaxation and cyclic circuit relaxation techniques were applied to reduce time consumption. Tsou-An Wu et al. proposed a divide-and-conquer method, using greedy algorithms to partition the quantum circuit into circuit blocks \cite{Wu2022}, applying solvers within the blocks to find feasible mappings, and employing heuristic methods to find swap gate insertion methods between circuit blocks. Although the time consumption was reduced in heuristic-solver combined approaches, the heuristic part made the methods difficult to find depth-optimal solutions.

In summary, as the most promising and practically applied method for quantum computers, solver-based quantum qubit mapping approaches have made substantial progress. However, while obtaining better solutions, the increased search time of solver-based methods has become a pressing issue. The state-of-the-art solver-based method, OLSQ2 \cite{Lin2023}, proposes an iterative approach to solve for the optimal depth and swap gate count, reducing the solution time by simplifying the variable counts and employing a more efficient method of encoding variables and constraints. Nevertheless, there is still room for improvement in this method. In OLSQ2, each global search targeting all candidate solutions consumes significant time, but only the two searches closest to the optimal solution provide practical benefits. Based on the above observations, it is efficient and reasonable to improve the solution efficiency by pruning the solver's search space.

In this work, we propose a machine learning-based approach for accelerating optimal qubit mapping (MLQM) using quantum circuit feature datasets. By constructing quantum circuit feature datasets based on quantum circuits and employing data augmentation and refinement techniques, MLQM aims to provide prior knowledge about the circuit depth and swap gate count using machine learning before the solver's optimization process and conducts the global search space pruning. The scheme allows the global search over all candidate mapping solutions to start from an initial state closer to the optimal circuit depth and swap gate count. Furthermore, MLQM optimizes the local search process within the solver for constraint verification of candidate mapping schemes, ultimately achieving a comprehensive improvement in the efficiency of quantum circuit mapping.

The main contributions of this work are as follows:

\begin{enumerate}
	
	\item{MLQM innovatively introduces a concise and efficient machine learning approach to the quantum circuit mapping problem, providing high-quality prior knowledge to the solver. With a tailored data augmentation method, MLQM significantly reduces the global search space and accelerates the solving process.}
	\item{MLQM proposes an adaptive adjustment mechanism for solver hyperparameters, which converts search trends into dynamic constraints, reduces the local search space and substantially enhances mapping efficiency.}
	\item{Extensive comparative experiments on widely used quantum computer architectures and circuits demonstrate that MLQM achieves an average speedup ratio of 1.79 in solving time and an average reduction of 22\% in memory footprint compared to state-of-the-art methods while ensuring optimal mapping quality. These results showcase MLQM's ability to expand the advantages of solver-based methods further and it's potential for solving quantum circuit optimization problems.}
	
\end{enumerate}

\section{Background}
In this section, we elucidate the concepts related to qubit mapping and solvers and the application of solver methods in qubit mapping.

\subsection{Qubit mapping}

Qubit mapping refers to mapping or routing qubits in a quantum program to physical qubits in a quantum computer, ensuring that they satisfy the connectivity constraints of the quantum computer, thus rendering the quantum circuit executable \cite{liTacklingQubitMapping2019}. The qubit mapping problem has two inputs: the first is the quantum circuit, and the second is the coupling graph, which represents the connectivity of the quantum computer. A coupling graph is a graph $(P, E)$, where $P$ is the set of physical qubits, and $E$ is a set consisting of tuples $(p_j, p_k)$, where $p_j$ and $p_k$ are two physical qubits, indicating the presence of edges between them. Only the pairs of physical qubits connected by edges in $E$ can support two-qubit gates. 

\begin{figure}[tb] %
	\begin{minipage}[b]{\columnwidth}
		\subfloat[Coupling graph of IBM QX2.]{\includegraphics[width=0.24\columnwidth]{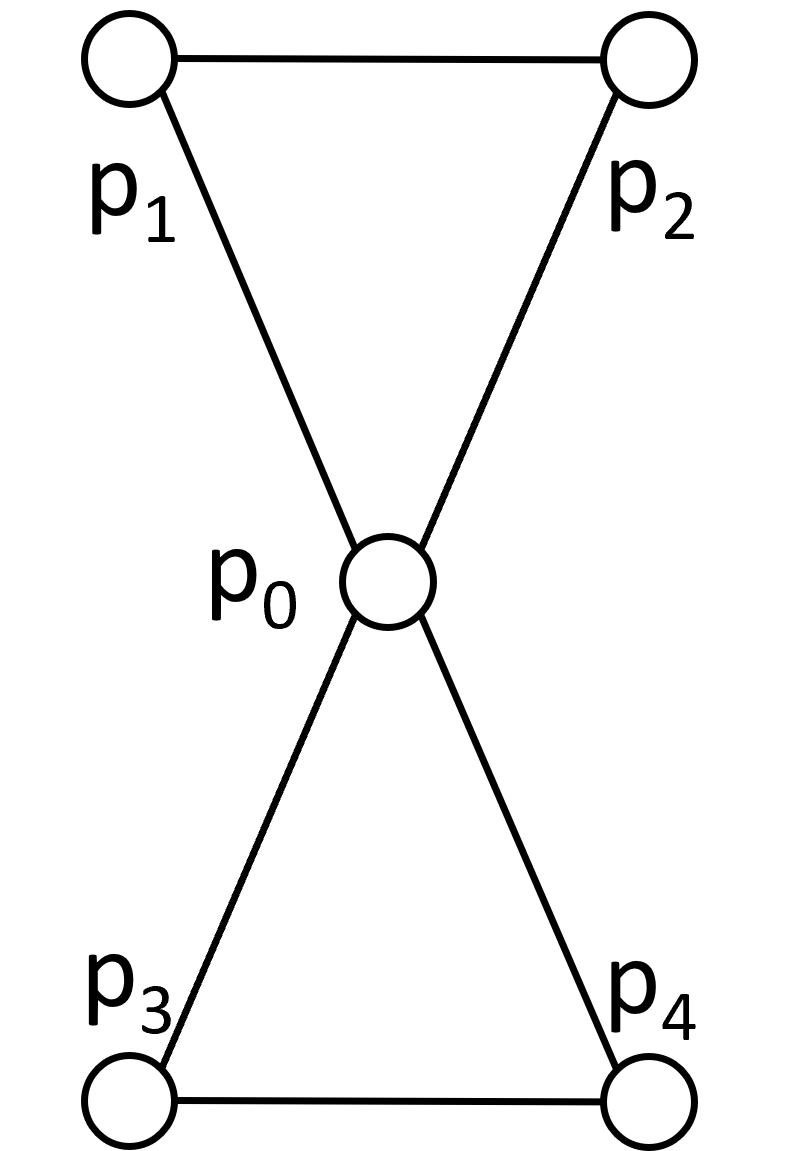}
			\label{qx}} 
		\subfloat[Quantum circuit before qubit mapping.]{\includegraphics[width=0.68\columnwidth]{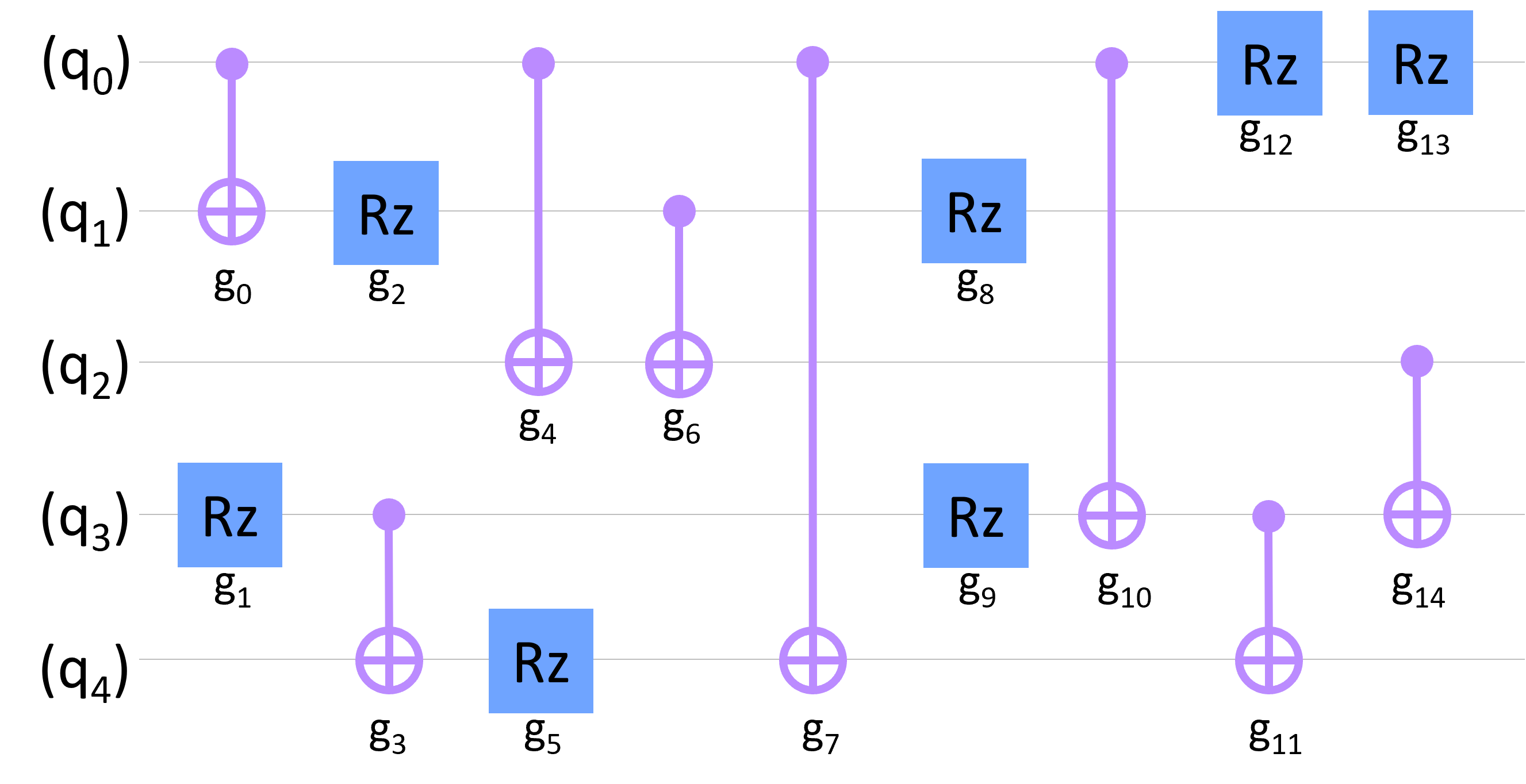} \label{qm_before}} 
		\\ 
		\subfloat[Result of qubit mapping.]{\includegraphics[width=\columnwidth]{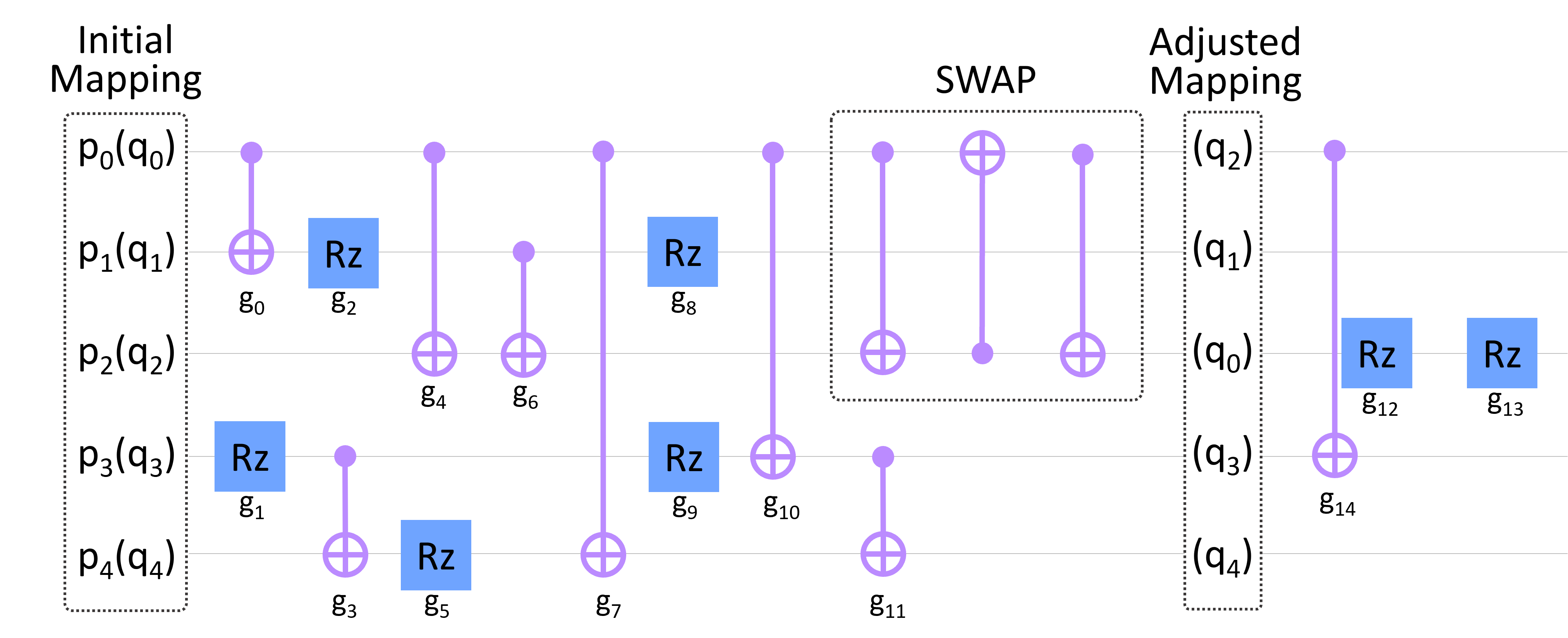} \label{qm_after}}
		\\
		\centering
		\subfloat[Dependency relationship of quantum gates in quantum circuit, the gate pointed to by any arrow must be executed after the gate at the start of the arrow.]{\includegraphics[width=0.8\columnwidth]{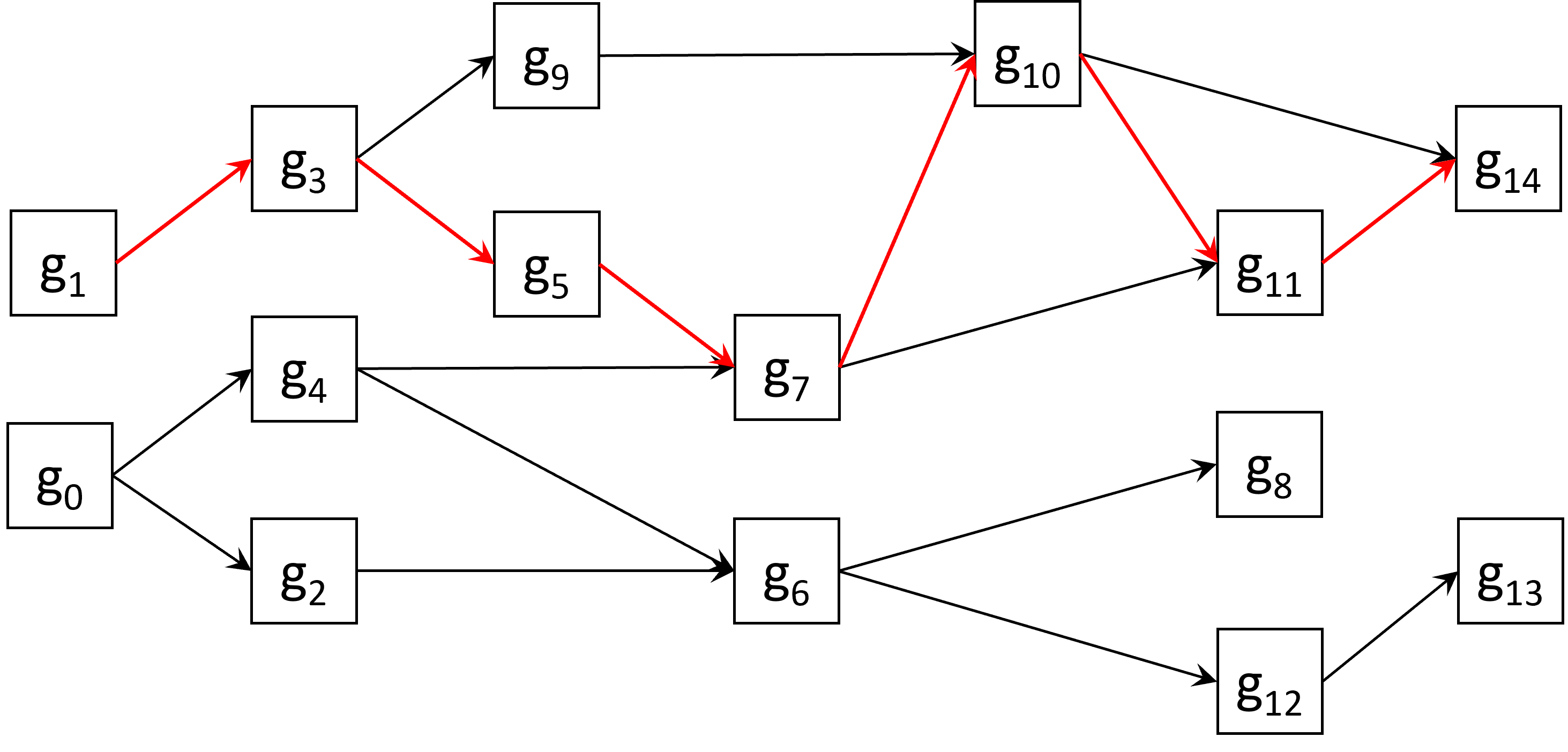} \label{qm_dag}}
	\end{minipage}
	\caption{The illustration of qubit mapping.}
	\label{qm}
\end{figure}

The process of qubits mapping requires particular attention to three essential tasks: proper initial mapping, connectivity compliance, and time compaction as follows:

\begin{enumerate}
	\item{\textit{Proper Initial Mapping}\\
		Quantum computer systems can be divided into a physical qubit layer and a logical qubit layer \cite{Gambetta2017}, representing quantum computer features and quantum circuit programs, respectively. Initial mapping refers to mapping logical qubits to physical qubits, enabling a quantum program to run on a quantum computer. Fig. \ref{qm}\subref{qm_before} illustrates the initial mapping process of logical qubits to physical qubits as $m_1:{q_0,q_1,q_2,q_3,q_4} \rightarrow {p_0,p_1,p_2,p_3,p_4}$.}\\
	
	\item{\textit{Connectivity Compliance}\\
		Connectivity Compliance involves inserting swap gates into the quantum circuit to satisfy the connectivity requirements of the quantum computer \cite{Linde2023}. As shown in Fig. \ref{qm}, a swap gate, composed of $3$ CNOT gates, can exchange the two logical qubits it operates on, thereby adjusting the mapping between logical and physical qubits. Fig. \ref{qm}\subref{qm_after} illustrates the routing process: gates $g_0$ to $g_{13}$ can operate properly because each acts on connected physical qubits or they are single-qubit gates, but $g_{14}$ acts on $p_2$ and $p_3$, violating the connectivity. Therefore, a swap gate is inserted on $q_0$ and $q_2$ to address this issue, adjusting the mapping to $m_1:{q_0,q_1,q_2,q_3,q_4}\ \rightarrow {p_2,p_1,p_0,p_3,p_4}$. After this, $g_{14}$ acts on the connected physical qubits $p_0$ and $p_3$. Connectivity compliance has a significant impact on the depth of quantum circuits.}\\
	
	\item{\textit{Time Compaction}\\
		Time Compaction focuses on adjusting the execution time of each quantum gate to make the quantum circuit depth as shallow as possible, and its process must satisfy the sequence of gates in the quantum circuit. As shown in Fig. \ref{qm}\subref{qm_dag}, $g_{12}$ and $g_{14}$ have no dependency constraints, so they can be arranged to execute in parallel rather than one after another, reducing the execution time of the circuit and increasing the utilization of qubits. The longest dependency chain (LDC) is another important concept in qubit mapping, marked in red in Fig. \ref{qm}\subref{qm_dag}.}\\
\end{enumerate}

It is worth noting that after qubit mapping, the circuit's length increased from 9 to 12 due to the additional insertion of swap gates. Solver-based methods can achieve a qubit mapping solution with optimal depth and the minimum number of swap gates to ensure circuit fidelity \cite{Lin2023}. However, more efficient solver-based schemes are necessary for quantum circuits with a larger number of qubits and greater depth to mitigate the computational complexity and resource requirements.

\subsection{Solver}
Solvers are tools for determining the satisfiability of formulas in SMT and Boolean Satisfiability (SAT) theories \cite{Moura2008}. In the context of qubit mapping tasks, solver methods can be employed to formulate the mapping problem as a constraint satisfaction or optimization problem. By encoding the constraints of the qubit mapping problem into a formulation the solver understands, it becomes possible to search for feasible mappings that satisfy the given constraints. Among them, the search for all mapping schemes of the current circuit can be regarded as a global search. In contrast, the search for mapping schemes under constraint conditions is considered a local search, both of which have a crucial impact on search efficiency.

The inputs of a solver typically comprise variables reflecting the values in the qubit mapping problem, along with a set of constraints and objectives relevant to the problem. The following variables are defined to match the qubit mapping problem:

\begin{enumerate}
	\item{The variable ${\pi_{q}^{t}}$ represents the set of physical qubits, which are mapped from logical qubits $q$ at time $t$, where the variable $t$ ranges from 0 to $T_{UB}$, an artificial set value that limits the execution time of all quantum gates.}
	
	\item{The swap gate execution time variables set is ${\sigma_{e}^{t}}$, where $e$ denotes the edge in the coupling graph and $t$ represents gate execution time, and the variable $t$ also ranges from 0 to $T_{UB}$.}
	
	\item{The set of time variables ${t_{g}}$ represents the execution times of quantum gates, with each element indicating the moment of execution of gate $g$.}
\end{enumerate}

All the above variables are of the same type as \cite{Lin2023}. Namely, variables in ${\pi_{q}^{t}}$ and ${t_{g}}$ are bit vectors, while variables in ${\sigma_{e}^{t}}$ are boolean variables. After the variable creation, the following constraints will be added to the solver:

\begin{enumerate}
	\item{\textbf{Mapping Constraint}: If $q\in Q$, then $\pi_{q}^{t}$ must equal to a physical qubit $p\in P$, where $Q$ and $P$ are the sets of all logical qubits and all physical qubits, respectively. Additionally, at the same time step, multiple logical qubits cannot be mapped to the same physical qubit.}
	\item{\textbf{Gate Execution Constraint}: The qubit pair of any two-qubit gate must be mapped to an edge in $E$ at the two-qubit gate execution time.}
	\item{\textbf{Execution Order Constraint}: The execution order of gates after mapping must adhere to the execution order of gates in the logical quantum circuit.}
	\item{\textbf{Swap Gate Constraint}: All swap gates must not have time overlap with other swap gates or logical quantum gates if they operate simultaneously.}
	\item{\textbf{Mapping Transformation Constraint}: If $q_{1}$ and $q_{2}$ are two logical qubits of a swap gate, then $\pi_{q_{1}}^{t}$ and $\pi_{q_{2}}^{t}$ will exchange their values after the completion time of the swap gate.}
\end{enumerate}

\subsection{Solve Optimal Qubit Mapping Problem with Solver}
The conventional methodology for attaining the optimal qubit mapping, as outlined in \cite{Lin2023}, begins by setting a specific depth limitation $T_B$ as an additional constraint. This limitation is then adjusted based on the solver's checking result until the optimal depth is reached. Once the optimal depth is determined, it becomes a fixed constraint, and the optimization goal shifts to minimizing the number of swap gates, denoted as $S_B$. The procedure for identifying the optimal number of swap gates closely mirrors the depth determination process. Ultimately, the solution that simultaneously satisfies both the optimal depth and the minimal number of swap gates under the optimal depth constraint can be considered the optimal qubit mapping.

\begin{figure}[tbp]
	\centering
	\includegraphics[width=1\columnwidth]{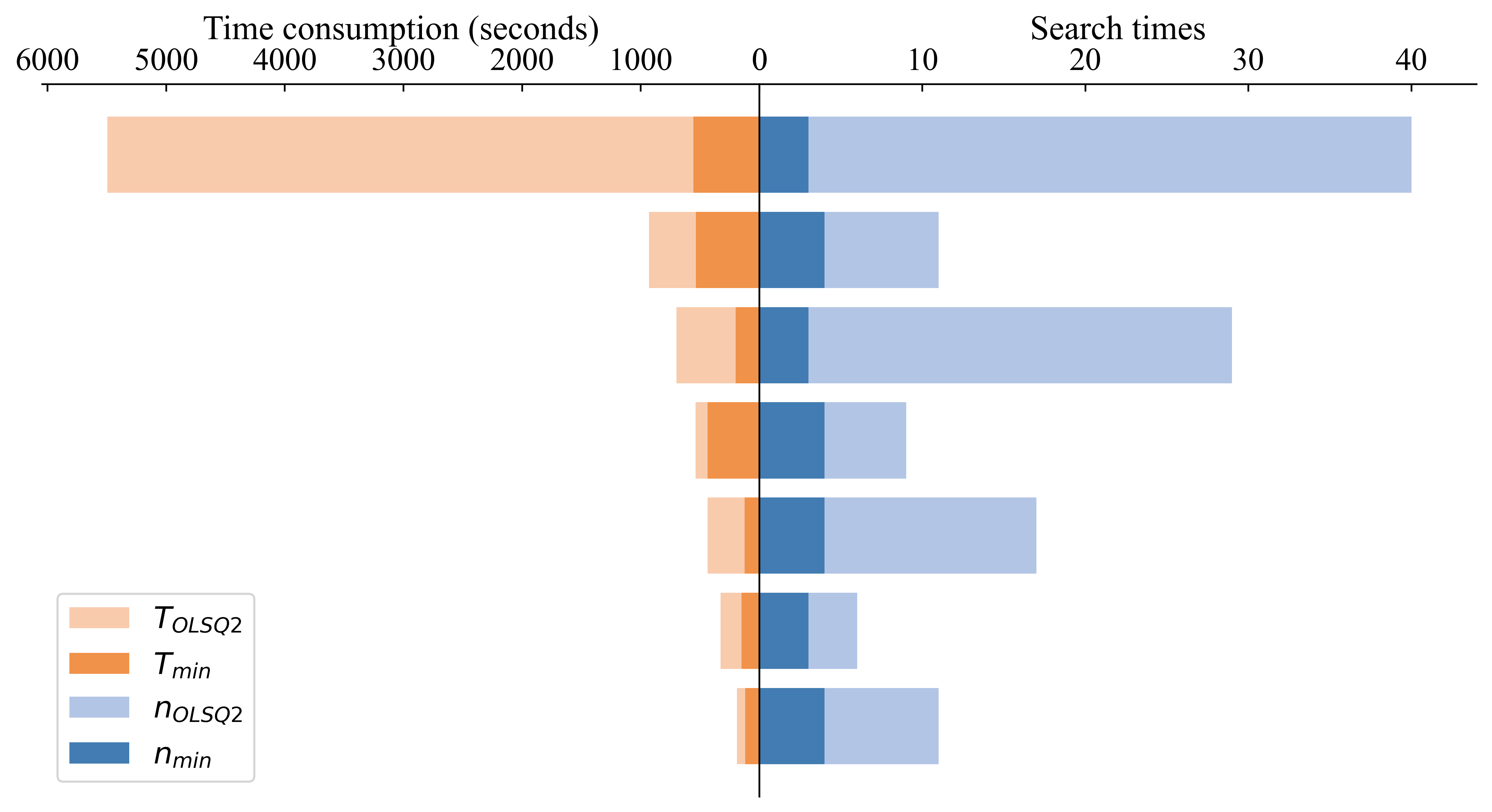}
	\caption{Time consumption and search counts of full OLSQ2 processes and only the necessary searches, $n_{\text{OLSQ2}}$ and $T_{\text{OLSQ2}}$ are the search count and total time consumption of OLSQ2, $n_{min}$ refers to minimal search count and $T_{min}$ is the time consumption of the searches only necessary.}
	\label{fig:olsq2_time}
\end{figure}

\section{Approach}
In this section, we will first present an analysis of the problem in current qubit mapping approaches, followed by an exposition of our proposed approach, which encompasses construction of quantum circuit feature dataset, model architecture, prediction, training procedures and solving process.

\subsection{Motivation}

The OLSQ2 algorithm is the state-of-the-art solver-based optimal qubit mapping algorithm \cite{Lin2023}, which proposes an iterative approach to solving for the optimal depth and swap number and reduces the solution time by simplifying the variable counts and employing an efficient method of encoding variables and constraints. However, the approach still has room for improvement. In OLSQ2, each search consumes a significant amount of time, but only the last two searches are valid because they determine the optimal solution. If the number of useless searches can be minimized, then the solution time will also be reduced significantly. Fig. \ref{fig:olsq2_time} runs OLSQ2 on seven circuits from the quantum benchmark dataset QASMBench \cite{Li2023} and demonstrates the benefits of reducing search counts. Comparing the time consumption between OLSQ2 and minimizing search counts, a maximum of 71.1\% and an average of 43.8\% reduction in time consumption can be achieved by minimizing the search count. Furthermore, qubit mapping depends solely on the execution order and type of quantum gates (single-qubit gate or two-qubit gate) of the quantum circuit and the coupling graph of quantum computers. Thus, the time required for extracting metrics from the quantum circuit is negligible compared to the time required in qubit mapping using solver-based methods, which means the prediction process is fast and real-time.

\begin{figure*}[tb]
	\centering
	\includegraphics[width=0.9\textwidth]{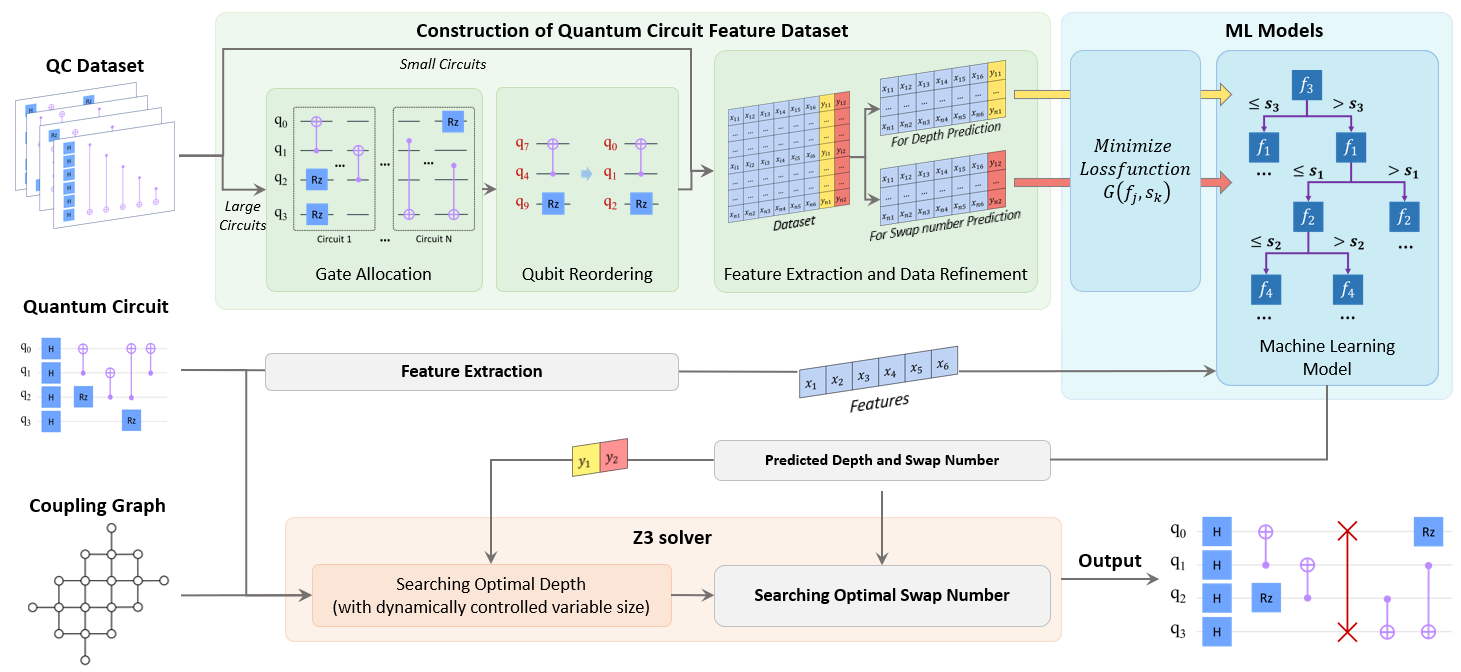}
	\caption{Overview of MLQM: Firstly, the quantum circuit dataset is enhanced by Gate Allocation and Qubit Reordering, which are then used to construct the quantum circuit feature dataset (green part). The machine learning model trained on this dataset predicts the depth of input quantum circuits and the number of swap gates (blue part). This prediction is used to reduce the global search space and accelerate the solving process. Additionally, the dynamically controlled variable size adopted in the solving process further reduces the local search space and enhances mapping efficiency (orange part).}
	\label{fullmodel}
\end{figure*}

Based on these observations, it is reasonable to predict the optimal qubit mapping using prior knowledge of the quantum circuit. The proposed qubit mapping method is illustrated in Fig. \ref{fullmodel}. By constructing datasets of quantum circuit features, we aim to predict the circuit depth and the number of swap gates beforehand. This approach allows us to start the search from an initial state closer to the optimal circuit depth and swap gate count, reducing the number of searches and ultimately improving efficiency.

\begin{algorithm} [tbp]
	\caption{Gate Allocation and Qubit Reordering} 
	\label{alg:dataset}
	\KwIn{$G=\{g_1,g_2,...\}$, $B=\{b_1,b_2,...\}$, $S=\{\}$, $D = \{\}$} 
	\KwOut{Quantum circuits set $D$ generated from $G$} 
	$num\_list = []$\;
	\For{$g_i$ in $G$} {
		Remove $g_i$ from $G$\;
		Add $g_i$ to $S$\;
		\If{length$(S)=b_j$ or \\
			\quad ($G\:is\:empty$ and $\exists\:two\:qubit\:gate\in S$)}
		{ 
			$j=j+1$\;
			\For{$g$ in $S$}{
				\For{$q$ in $g$}{
					\If{$q\notin num\_list$}{Append $q$ to $num\_list$;}
				}
			}
			\For{$g$ in $S$}{
				\For{$q$ in $g$}{
					\If{$\exists$ $num\_list[i]$ equals $q$}{$q \leftarrow i$;}
				}
			}
			Save $S$ to $D$;
		} 
	}
	return $D$\;
\end{algorithm}

\subsection{Construction of the Quantum Circuit Feature Dataset}
Our quantum circuit feature dataset is constructed based on the QASMBench dataset provided in \cite{Li2023}. This dataset integrates quantum circuits commonly used in chemistry, simulation, search, optimization, and machine learning. Due to noise and gate imperfections, quantum computers can only reach a finite depth before the advent of total decoherence \cite{Fauseweh2024}, rendering excessively long quantum circuits impractical. Therefore, we choose the quantum circuits in the QASMBench dataset whose qasm file's size is less than 2KB for feature extraction and labeling. Unfortunately, the number of quantum circuits meeting this requirement in the QASMBench dataset does not exceed 50, making it nearly impossible to train a model with sufficient accuracy on such a small dataset. To address this issue, we propose a method for enhancing quantum circuit datasets from circuits that do not meet the length requirement. This method involves two steps: gate allocation and qubit reordering.

In gate allocation, for every large quantum circuit $G$ in the quantum circuit dataset, we sequentially search its quantum gates and append the gates to an empty set $S$ one by one until the number of elements in the set reaches an upper bound $b$, then $S$ is saved to $D$, all elements in $D$ will proceed to qubit reordering. It's worth noting that $b$ is a dynamic value in a set $B$, which dramatically enhances the diversity of circuits. Additionally, splitting quantum circuits by size $b$ may lead to residuals in the circuit since the number of gates in a quantum circuit may not necessarily be the sum of dynamic values $b$ in $B$. For each residual, if it contains at least one two-qubit gate, it will be recorded in $D$. In the qubit reordering section, the qubits of each quantum circuit in $D$ are renumbered while preserving the gate dependency relationships, ensuring the circuits are adapted for subsequent processing. The steps for dataset enhancement are shown in Algorithm \ref{alg:dataset}.

To comprehensively describe the properties of quantum circuits in the context of the qubit mapping problem, circuit depth and circuit width are included in the features as the most intuitive and fundamental characteristics of quantum circuits first. Additionally, max qubit depth and operation density are added since they provide insights into how gates are spread across the quantum circuit. Furthermore, the two-qubit gate count and entanglement variance are incorporated into the feature set, as the arrangement of two-qubit gates is central to the qubit mapping problem. A more detailed explanation of these features and their relation to the qubit mapping problem is as follows:

\begin{enumerate}
	\item{\textbf{Circuit depth}: This metric represents the length of the longest dependency chain(LDC) in the quantum circuit, given by (\ref{eq_cd}), an example of LDC is shown in Fig. \ref{qm}\subref{qm_dag}. LDC determines the lower bound of the circuit depth after qubit mapping.
	}
	\begin{equation}
		\label{eq_cd}
		\text{Circuit depth}=length(LDC).
	\end{equation}
	\item{\textbf{Circuit width}: This metric refers to the number of qubits present in the quantum circuit and defined in  (\ref{eq_cw}).
	}
	\begin{equation}
		\label{eq_cw}
		\text{Circuit width}=n_{qubits}.
	\end{equation}
	\item{\textbf{Max qubit depth}: Max qubit depth is obtained by counting the total number of gates on each qubit and selecting the maximum among them, shown in (\ref{eq_mqd}), where $n_{q,gates}$ represents the number of quantum gates on different qubits $q$.}
	\begin{equation}
		\label{eq_mqd}
		\text{Max qubit depth}=\max(n_{q,gates}).
	\end{equation}
	\item{\textbf{Operation density}: Which is also known as gate density, refers to the occupancy rate of quantum gates in both the time and space dimensions of the qubits. A higher operation density will increase the circuit depth and the number of required swaps after qubit mapping. The formula for calculating the operation density is given by (\ref{eq_od}), where $n_{1}$ denotes the count of single-qubit gates, and $n_{2}$ represents the count of two-qubit gates in the quantum circuit.}
	\begin{equation}
		\label{eq_od}
		\text{Operation density}=\frac{n_{1}+2\times n_{2}}{length(LDC)\times n_{qubits}} .
	\end{equation}
	\item{\textbf{Two-qubit gate number}: The two-qubit gate number is defined as the count of two-qubit gates present in the quantum circuit. Typically, an increase in the number of two-qubit gates will lead to higher circuit depth and a more significant number of required swaps after mapping when other parameters remain constant:}
	\begin{equation}
		\label{eq_tqgn}
		\text{Two-qubit gate number}=n_{\text{two-qubit gates}}.
	\end{equation}
	\item{\textbf{Entanglement variance}: Entanglement variance describes the uneven distribution of two-qubit gates across different qubits. Qubit mappings with larger entanglement variance will benefit more from the qubit mapping process. This concept is captured by (\ref{eq_ev}), where $n_{2,q}$  represents the number of two-qubit quantum gates on qubit $q$, while $\overline{n_{2}}$ denotes the average number of two-qubit quantum gates across all qubits.}
	\begin{equation}
		\label{eq_ev}
		\text{Entanglement variance}=\frac{\ln_{}{(\sum (n_{2,q}-\overline{n_{2}} )^2 +1)}}{n_{qubits}} .
	\end{equation}
\end{enumerate}

 Further exploration of features is demonstrated in \ref{app:regression tree}. After feature extraction, the optimal depth and the minimum number of swap gates are calculated for each sample using OLSQ2. Subsequently, based on the result and the existing features, we constructed two quantum circuit feature datasets, taking the depth and the number of swap gates of quantum circuits as their labels, respectively. The feature datasets are available after the data refinement process, which uses the AllKNN algorithm to process each sample iteratively\cite{1976}. The algorithm retains each sample by examining the types of its $n$ nearest neighbors, with $n$ incrementally increasing from 1. This approach mitigates data imbalance, enhancing the datasets' suitability for predicting depth and swap gate numbers. The constructed datasets are detailed in \ref{app:mlqd}.
 
\subsection{Model Training and Prediction}
Considering the relatively small size of our dataset and the interpretability and training efficiency of the model, we employ a tree regression model for label prediction. The following loss function $G$ is minimized to determine the feature and its split point at each step of the tree partitioning process:

\begin{equation}
	\label{loss1}
	G(f_{j},s_{k})=\frac{n_{l}(f_{j}, s_{k})L_{l}(f_{j},s_{k})}{n}+\frac{n_{r}(f_{j},s_{k})L_{r}(f_{j},s_{k})}{n}.
\end{equation}
\begin{equation}
	\label{loss2}
	L_{d}(f_j,s_k)=\frac{1}{n_d}\sum_{y\in Q_d}^{}(y-\overline{y_d})^2.
\end{equation}

Here, $n$ is the number of samples, $f_j$ represents the feature being split, $s_k$ is the value used in feature splitting, $n_l$ is the number of samples less than $s_k$, and $n_r$ refers to the number of samples greater than $s_k$ after $f_j$ is split. $Q_d\ (d=l\ or\ r)$ represents the feature values less than or equal to and greater than $s_k$ after $f_j$ is split. During the model training process, the algorithm attempts to find the splitting method that minimizes the loss function. After training, the predicted depth $d_{pred}$ and the predicted swap gate number $n_{pred}$ of a quantum circuit are obtained. Examples of tree regression models for circuit depth and swap number prediction are presented in \ref{app:regression tree}.

\subsection{Solving}

\begin{algorithm} [tbp]
	\caption{Dynamically Adjusting Method of Variable Size in Solving} 
	\label{alg:update}	
	\KwIn{Solver, Satisfiability, $d_{try,s}$, $d_{ub}$, $l_b$, $\{\sigma_{e}^{t}\}$,  $\{\pi_{q}^{t}\} $, $\{t_{g}\}$} 
	\KwOut{Updated solver, $\{\sigma_{e}^{t}\}$,  $\{\pi_{q}^{t}\} $ and $\{t_{g}\}$} 
	\uIf{Satisfiability == Unsatisfied}
	{\If{$bits(d_{try,s})\textgreater bits(d_{try,s-1})$ or $d_{try,s}+2\textgreater d_{ub}$}{
			\uIf{$d_{try,s}+2 \textgreater d_{ub}$ and $d_{try,s}\textless d_{th}$}
			{ $d_{ub}\:=\:d_{try,s}+d_{dl}$;}
			\ElseIf {$d_{try,s}+2 \textgreater d_{ub}$ and $d_{try,s}\textgreater d_{th}$}{
				$d_{ub}\:=\:d_{try,s}+d_{ds}$;
			}
			\If{$bits(d_{try,s})\textgreater bits(d_{try,s-1})$}{
				$l_b$ = $\lfloor log_2(d_{try,s}) \rfloor + 1$;}
			Update $\{\pi_{q}^{t}\} $ with $d_{ub}$;\\
			Update $\{t_{g}\}$ with $l_b$;\\
			Update $\{\sigma_{e}^{t}\}$ with $d_{ub}$;\\
			Update solver's constraints with $d_{ub}$;\\
		}
	}
	\Else{
		\If{$bits(d_{try,s})\textless bits(d_{try,s-1})$}
		{$l_b$ = $\lfloor log_2(d_{try,s}) \rfloor + 1$;\\
			Update $\{t_{g}\}$ with $l_b$;\\
			Update solver's constraints with $d_{ub}$;}
	}
\end{algorithm}

\begin{figure}[tbp]
	\centering
	\includegraphics[width=1\columnwidth]{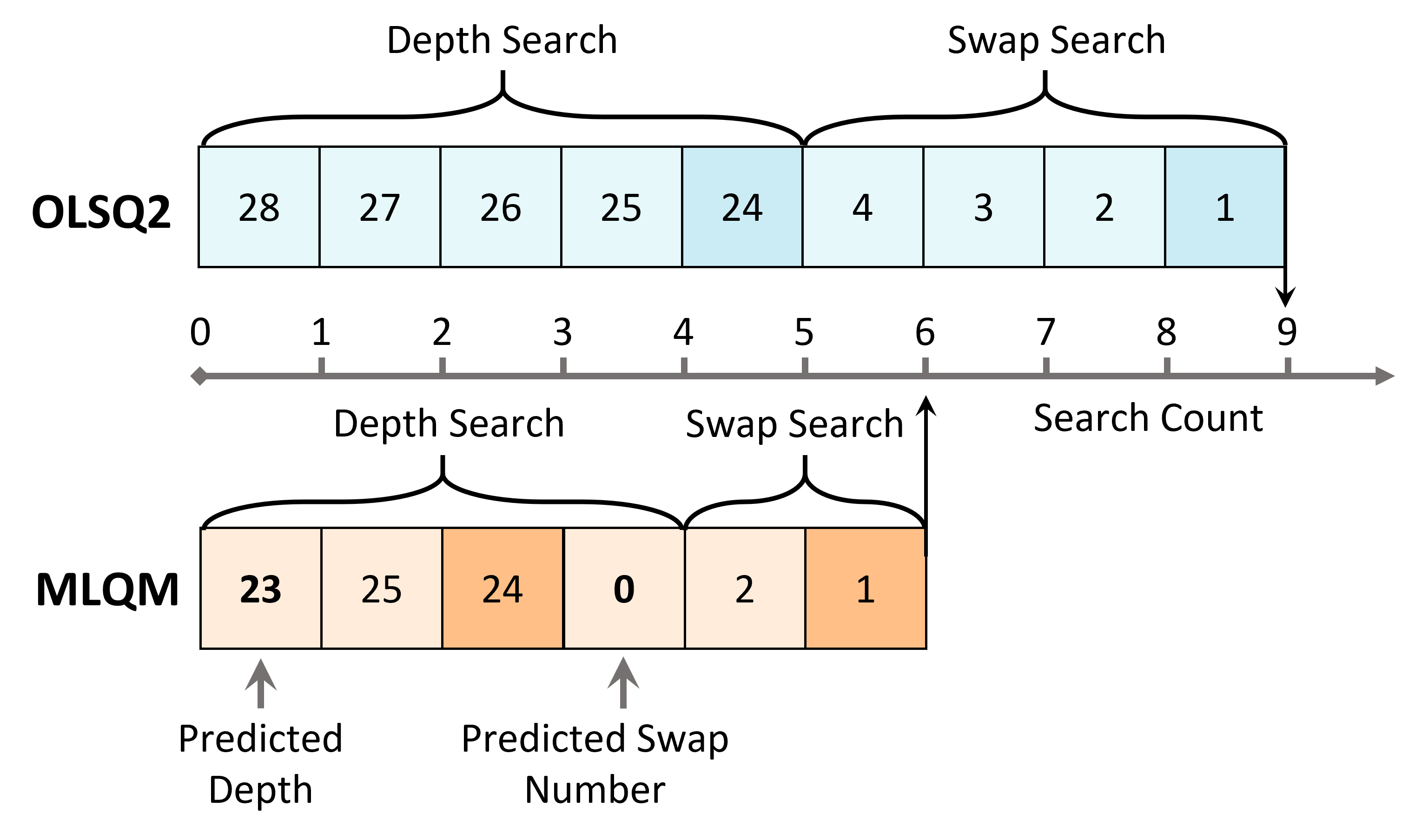}
	\caption{An example showing the difference between the solution processes of MLQM and OLSQ2, the blue sequence refers to the search process of OLSQ2 while MLQM's search process is painted in orange.}
	\label{fig:mlqm_vs_classical}
\end{figure}

The solver is iteratively utilized to find the optimal qubit mapping with Microsoft's Z3 solver\cite{Moura2008}. Considering the optimal circuit depth after qubit mapping cannot be less than the length of the LDC, the initial value of the depth searching is initialized to $max(d_{pred}, length(LDC))$. The swap number search initializes at $min(n_{pred}, n_{last})$ to ensure the selected swap number does not exceed the current optimal value, with $n_{last}$ representing the swap number from the previous satisfactory formulation in the search process. Subsequently, the search program incrementally adjusts these values in 2-unit steps to balance computational efficiency and solution accuracy. Once a transition is detected in the formulation's satisfiability, indicating the search has reached a critical boundary, the program determines the optimal solution by meticulously verifying the intermediate values between depth constraints. As illustrated in Fig. \ref{fig:mlqm_vs_classical}, during a depth searching process where the formulation proves unsatisfiable at depth 23 and satisfiable at depth 25, the algorithm sets the depth constraint to 24 for subsequent verification. If the formulation remains satisfiable, 24 is confirmed as the optimal depth; otherwise, 25 is retained .Similarly, the swap gate search process employs an identical optimization strategy as the depth search process. Fig. \ref{fig:mlqm_vs_classical} also demonstrates the search reduction of MLQM, wherein depth search iterations are diminished by 1 and swap search iterations are diminished by 2 relative to OLSQ2.

Another improvement to our article is derived from the following observation: in experiments, we note that the solving efficiency of the solver decreases with an increase of two factors: the number of variables and the number of bits in the bit vector variables. In \cite{Lin2023}, the size of the second dimension of $\{\sigma_{e}^{t}\}$ and $\{\pi_{q}^{t}\}$ and the bit number of variables in $\{t_{g}\}$ are according to 1.5 times the length of LDC. However, the method does not consider the asynchrony between $\{\sigma_{e}^{t}\}$, $\{\pi_{q}^{t}\}$ and $\{t_{g}\}$, and the reduction in search space as the circuit depth decreases.

We propose a method that increases the solving efficiency by pruning the local search space. More specifically, the method finely and dynamically adjusts the size of the time dimension of $\{\sigma_{e}^{t}\}$ and $\{\pi_{q}^{t}\}$, as well as the size of the bit vectors in $\{t_{g}\}$. First, we define six variables: $d_{try,s}$ as the depth of the circuit verified in step $s$, $l_b$ as the number of binary digits of every variable's size in $\{t_{g}\}$, $d_{ub}$ as the size of the time dimension of $\{\sigma_{e}^{t}\}$ and $\{\pi_{q}^{t}\}$, which must be larger than $d_{try,s}$, $d_{th}$ as the threshold value for circuit depth, and the last two variable $d_{dl}$ and $d_{ds}$ are step size of $d_{ub}$ .

After each solving process, the program will make a judgment based on the satisfiability of the formation. As algorithm \ref{alg:update} demonstrated, the solver's constraints will be adjusted under either of the two following situations. The first situation is $d_{try,s}$ continuing to grow and reaching $d_{ub}-2$, or its binary value exceeds $d_{try,s-1}$. If this situation is triggered by the first condition, $d_{ub}$ will be adjusted based on its value relative to $d_{th}$. If $d_{ub}$ is larger than $d_{th}$, $d_{ub}$ will be set to $d_{ub} + d_{dl}$, otherwise, it will be set to $d_{ub} + d_{ds}$, where $d_{dl} > d_{ds}$. The strategy is based on an observation: circuits with smaller depth usually require fewer search iterations to find the optimal solution. Similarly, $l_b$ will also be adjusted if $bits(d_{try,s})\textgreater bits(d_{try,s-1})$. After that, variables $\{\pi_{q}^{t}\}$ and $\{\sigma_{e}^{t}\}$ will be updated by variable $d_{ub}$, $\{t_{g}\}$ will be updated by $l_b$, and the solver's constraints will also be re-added. Another situation is the number of binary digits of $d_{try,s}$ is less than the number of binary digits of $d_{try,s-1}$. Then, $l_b$, $\{t_{g}\}$ and the solver's constraints will also be updated to accommodate the value of $d_{try,s}$. Ultimately, leveraging the MLQM approach enables the derivation of the optimal qubit mapping solution after depth searching and swap gate number searching, and the solving efficiency is greatly improved.

\textbf{\begin{figure*}[htbp] %
		\centering
		\subfloat[Google Sycamore.]{\includegraphics[width=0.2\textwidth]{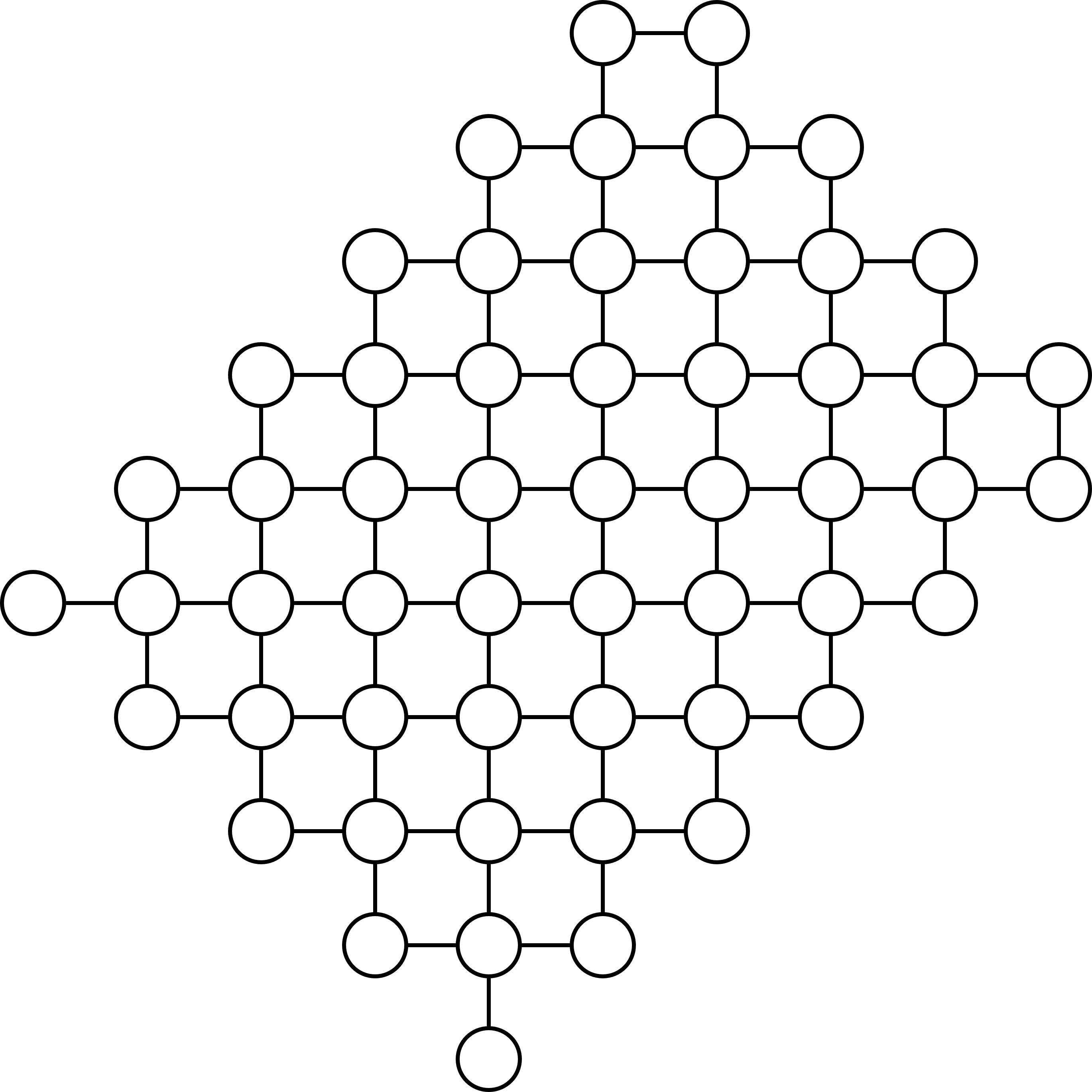}\label{sycamore}}
		\hspace{0.01\textwidth}
		\subfloat[IBM-Rochester.]{\includegraphics[width=0.18\textwidth]{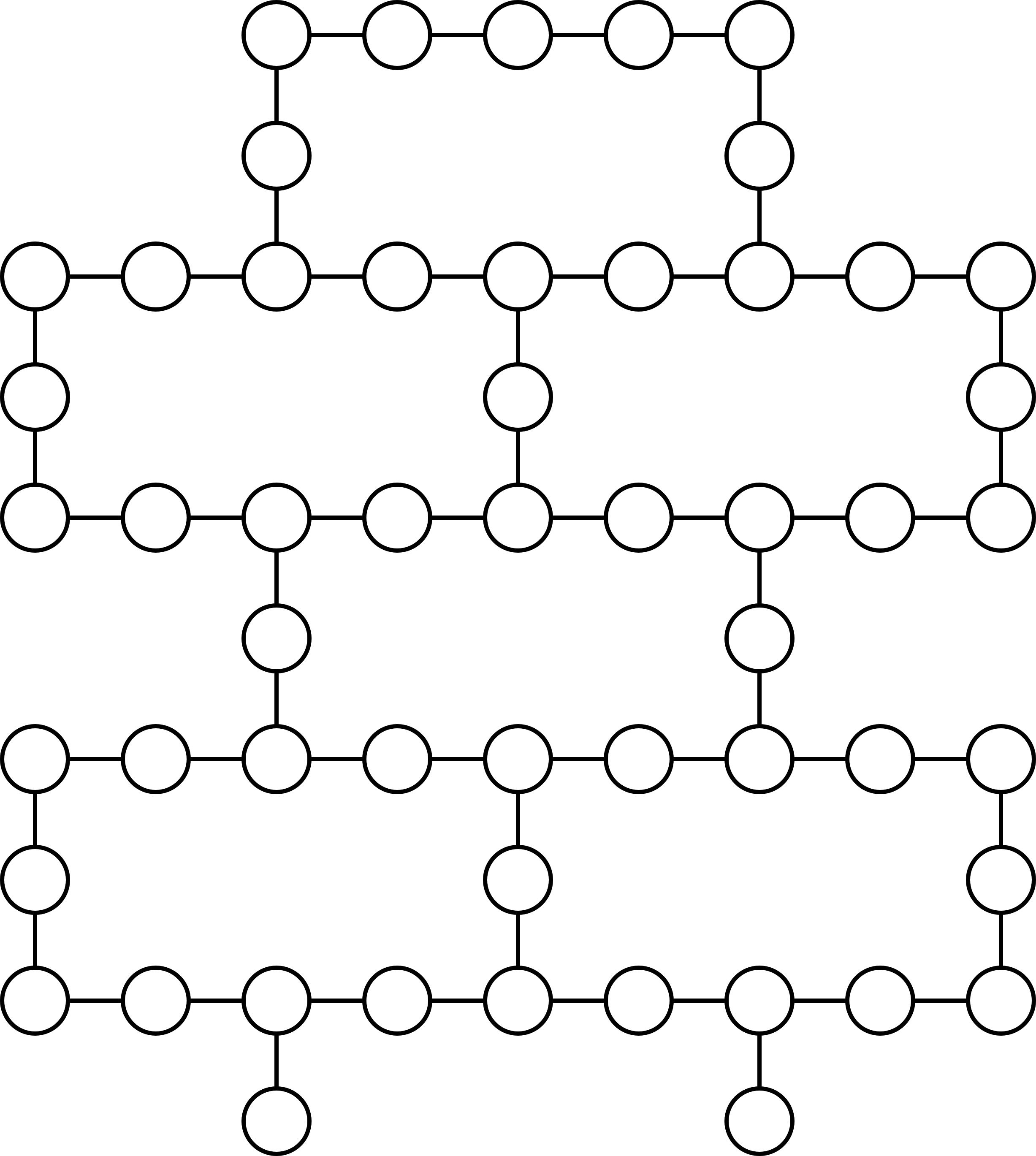}\label{ibmRochester}}
		\hspace{0.01\textwidth}
		\subfloat[Rigetti Aspen-4.]{\includegraphics[width=0.18\textwidth]{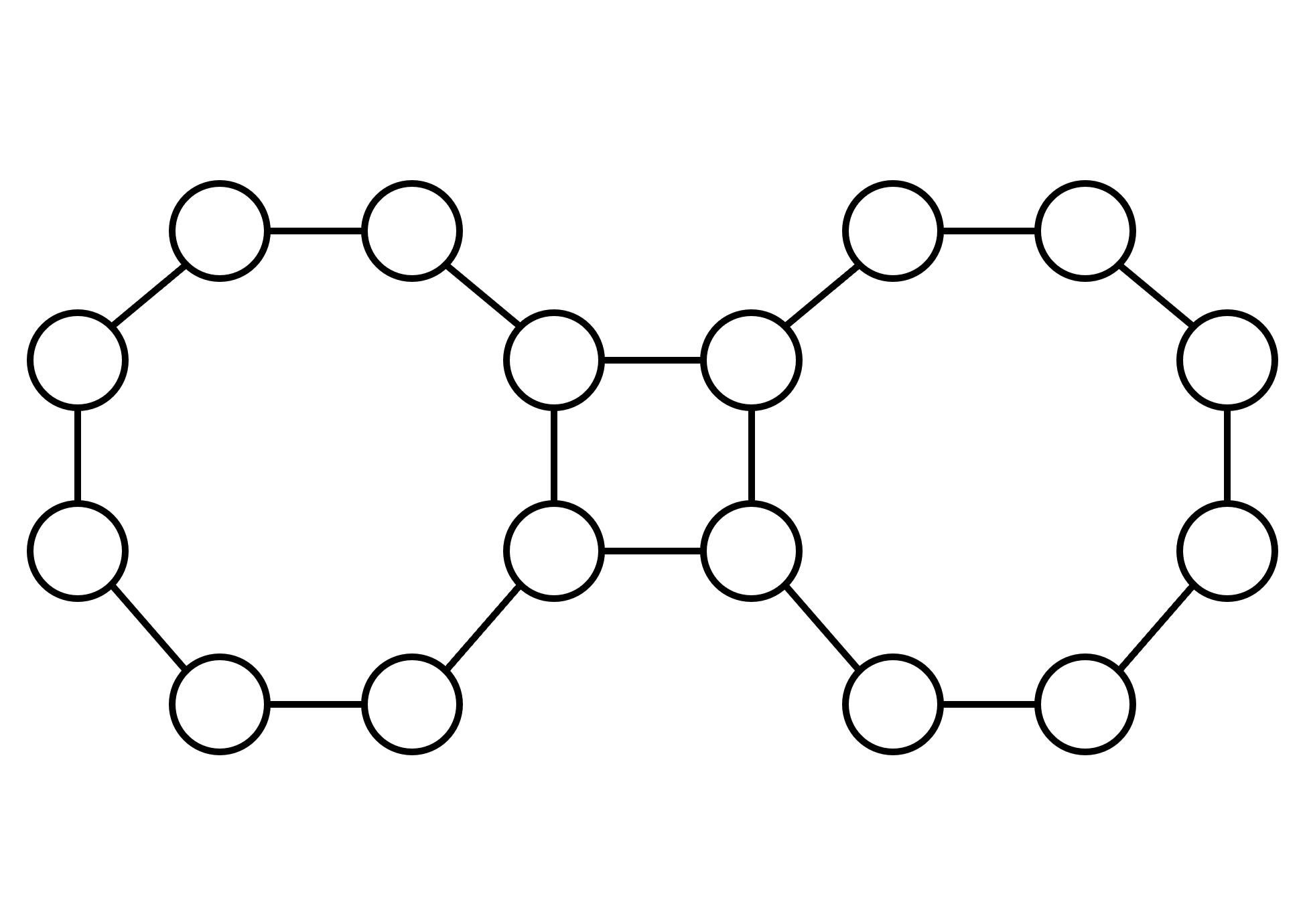} \label{aspen4}}
		\hspace{0.01\textwidth}
		\subfloat[IBM-Melbourne.]{\includegraphics[width=0.18\textwidth]{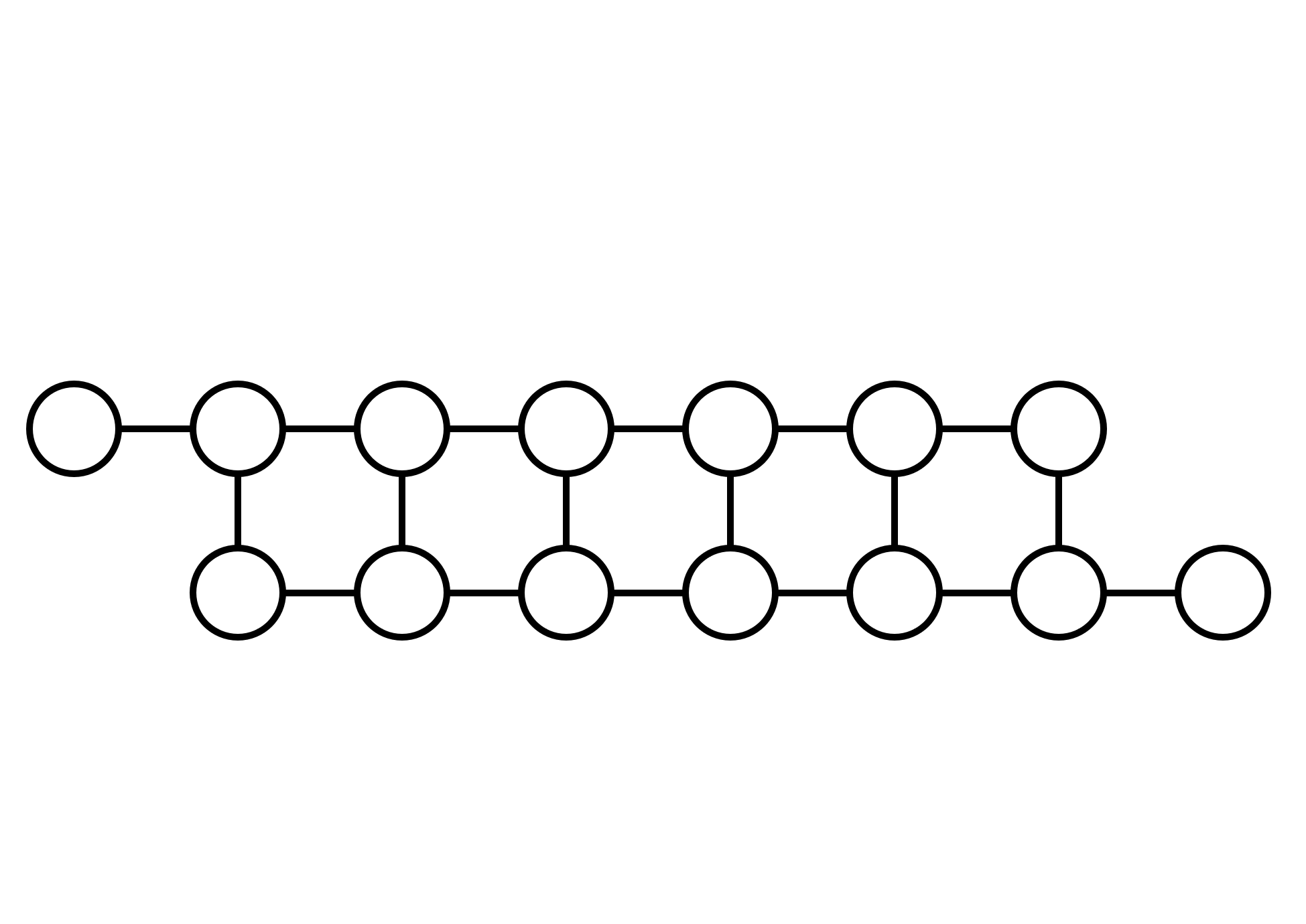} \label{ibmMelbourne}}
		\hspace{0.01\textwidth}
		\subfloat[5x5 grid.]{\includegraphics[width=0.13\textwidth]{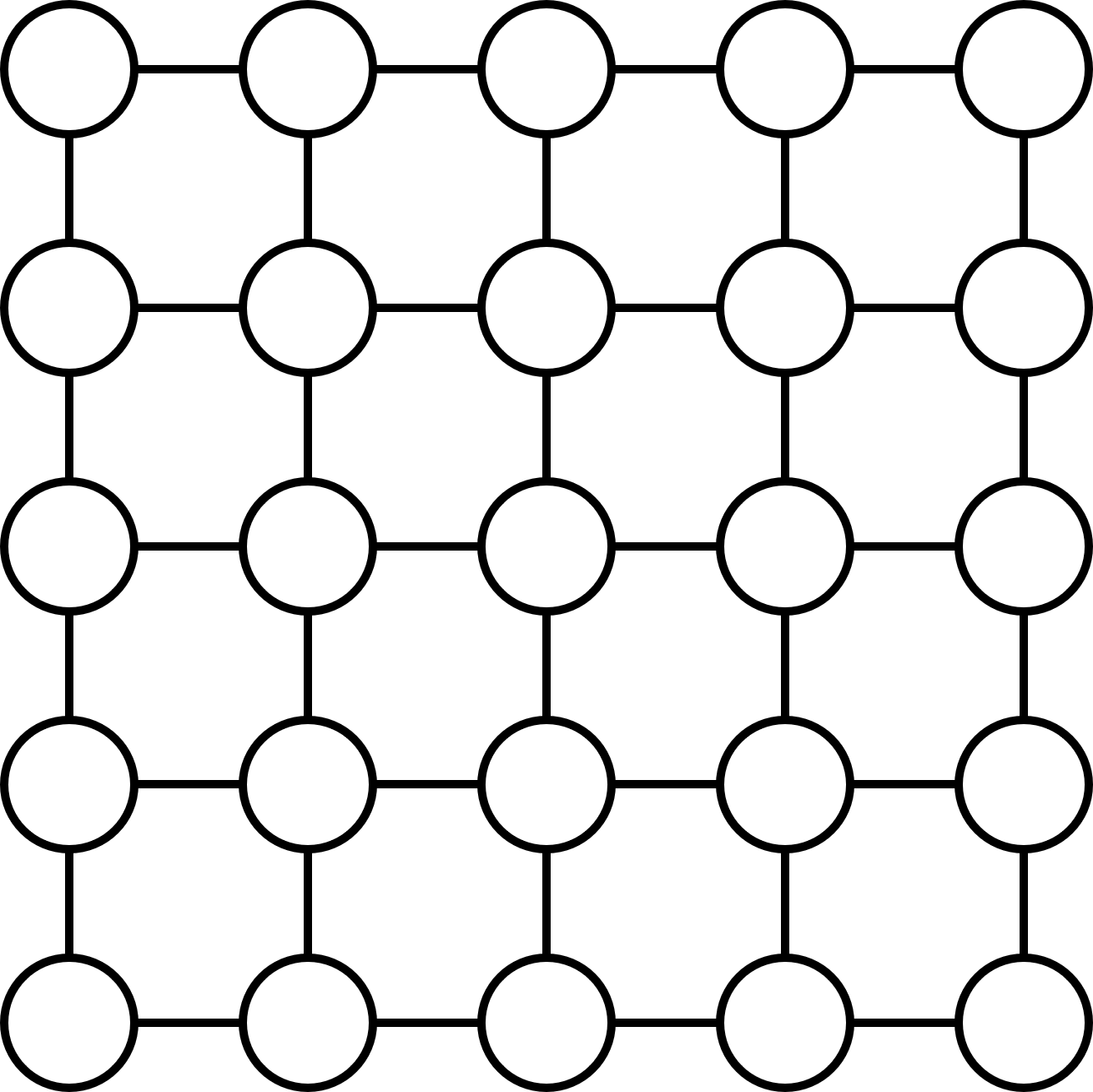} \label{5x5}}
		\caption{Coupling graphs used in the experiment, The circles in the figure indicate physical qubits, and the connecting lines between the circles indicate that two-qubit gates can be executed between physical qubits. Sycamore, IBM-Rochester, Aspen-4, IBM-Melbourne and 5x5 grid have 54, 53, 16, 14, 25 qubits respectively.}
		\label{exp_coupling}
	\end{figure*}
}

\section{Experimental Results}
\subsection{Experimental Settings}
In this section, we demonstrate the performance of the proposed solution relative to other baselines. Our method runs in a Python 3.6 environment, utilize the Z3 solver's Python API (v4.13.0.0) for constraint solving and the python-SAT library (v0.1.8.dev12) for constraint adding. We employ scikit-learn's Python API (v0.24.2) and imbalanced-learn (v0.8.1) for model training and dataset refinement. All experiments are conducted on an AMD Ryzen7 5700G CPU running at 3.8GHz with 32GB of RAM. 

We design two experiments to validate our method's efficiency and optimality. The first part compares our method's solving time and memory consumption and the latest solver-based approach, OLSQ2\cite{Lin2023}. The second part compares our method and the leading representative heuristic method, SABRE\cite{liTacklingQubitMapping2019}.

In constructing the quantum circuit feature dataset, the QASMBench dataset is enhanced through the utilization of a method of generating quantum circuit datasets twice, which expands and ensures diversity within the dataset. All single-qubit and two-qubit gates are retained in the first utilization, with a dynamic parameter $b$ less than 100. In the second utilization, only two-qubit gates are maintained, with $b$ less than 30. To avoid overfitting, the max depth of the tree regression model is set to 5.

The training set is built from the remaining portion of the quantum circuit feature dataset after the test set is removed. To comprehensively validate our approach, we perform experiments using five coupling graphs: Google's Sycamore with 54 qubits, IBM's Rochester and Melbourne with 53 and 14 qubits respectively, Rigetti's Aspen-4 with 16 qubits, and a 5x5 grid coupling graph with 25 qubits, as shown in Fig. \ref{exp_coupling}. All coupling graphs above are commonly used in qubit mapping verification. For the test set, considering the limited size of the original dataset, eight quantum circuits in QASMbench's original dataset are used, with qubit numbers ranging from 3 to 40 and qubit gates ranging from 25 to 107. The number in every sample's name represents the logical qubit numbers, for example, sample ``bv\_n14" has 14 qubits. For Sycamore and Rochester, all eight samples are used as test set. For Aspen-4, IBM-Melbourne and 5x5 grid, due to their qubit numbers, the samples in the test set with fewer qubits than their qubit counts are used. 
In the solving process, $d_{th}$, $d_{dl}$ and $d_{ds}$ are set to 50, 15, and 10 respectively to maintain solving efficiency and avoid unnecessary re-adding of variables and constraints.

\begin{figure}[tbp] %
	\begin{minipage}[b]{\columnwidth}
		\centering
		\subfloat[Comparison of time consumption on Sycamore. For consistency in plotting, the value of the samples marked with an asterisk (*) has been divided by 3.]{\includegraphics[width=0.45\columnwidth]{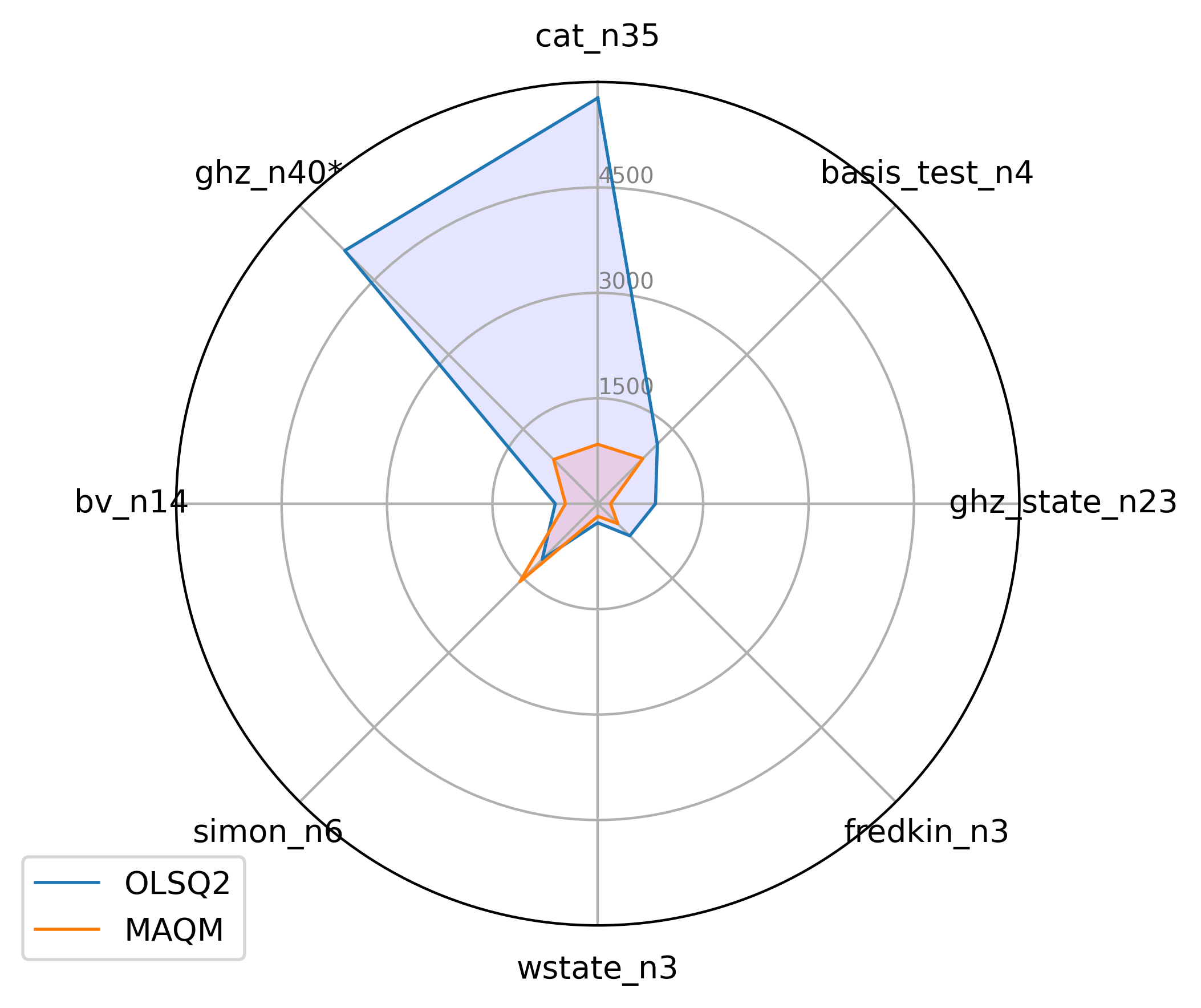}\label{syca_time}}
		\hspace{0.005\textwidth}
		\medskip
		\hspace{0.005\textwidth}
		\subfloat[Comparison of time consumption on IBM-Rochester. For consistency in plotting, the value of the samples marked with an asterisk (*) has been divided by 5.]{\includegraphics[width=0.45\columnwidth]{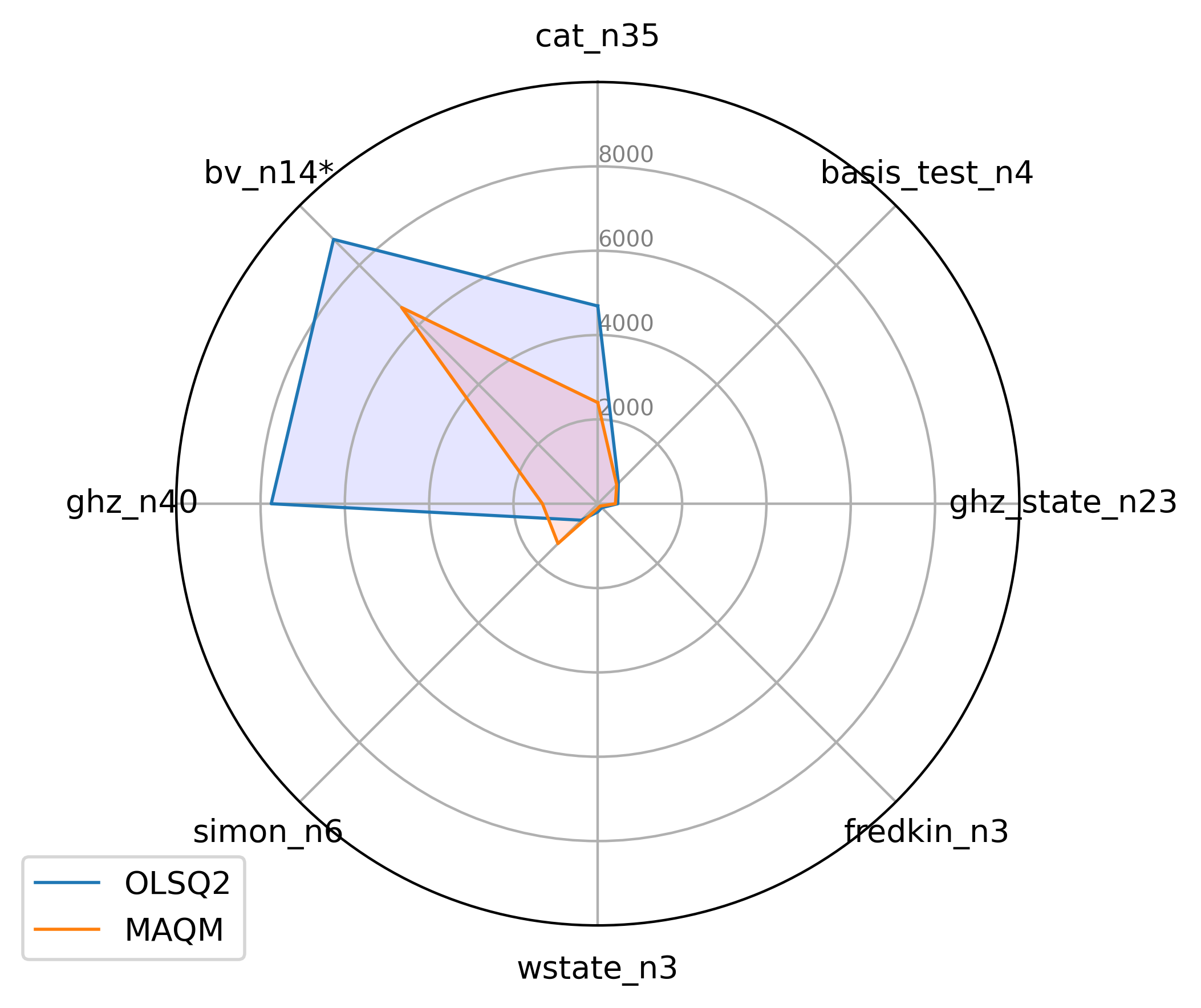}\label{ibmRochester_time}}
	\end{minipage}
	\\
	\begin{minipage}[b]{\columnwidth}
		\centering
		\subfloat[Comparison of time consumption on Aspen-4.]{\includegraphics[width=0.45\columnwidth]{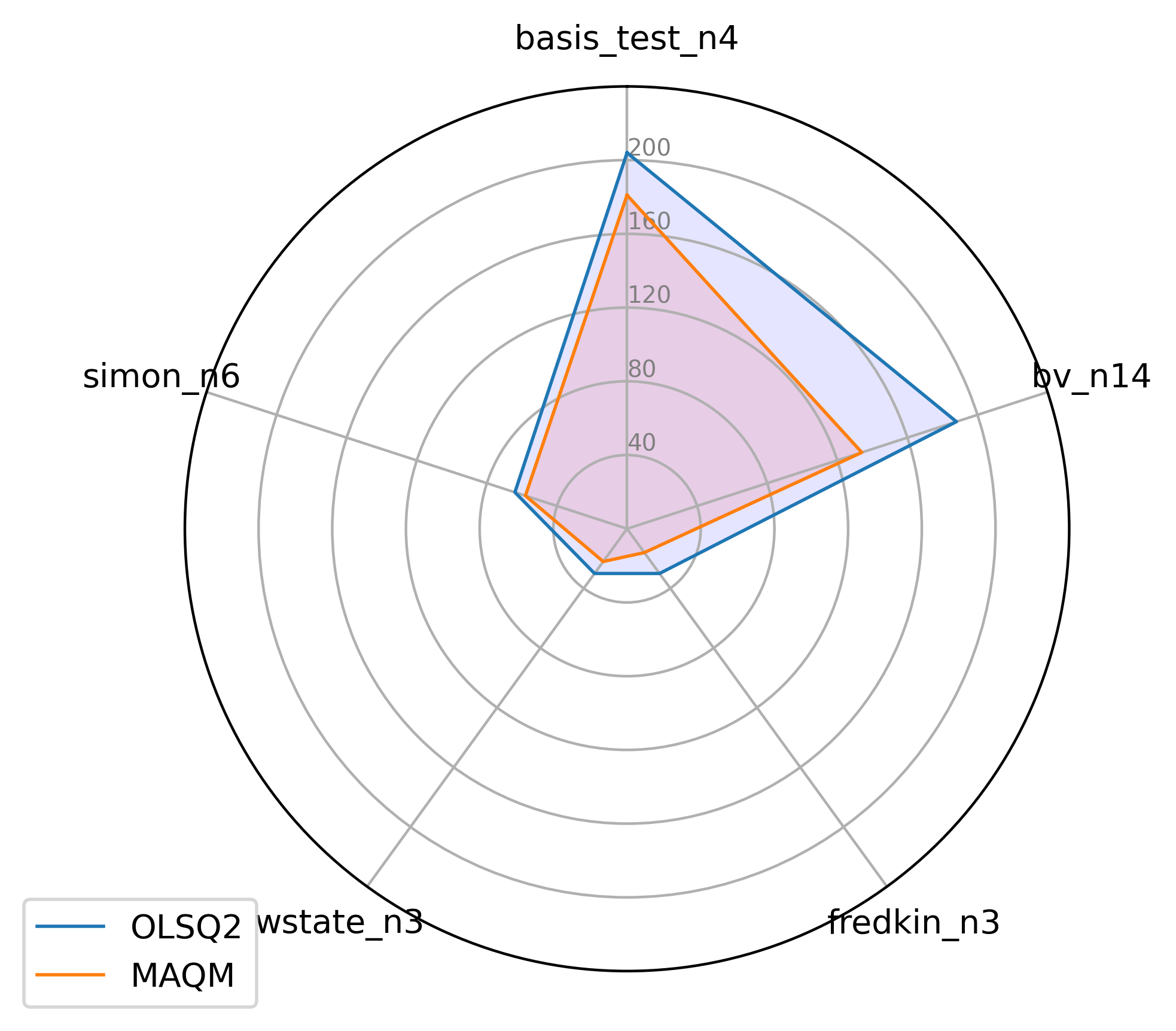} \label{aspen4_time}} 		
		\hspace{0.005\textwidth}
		\medskip
		\hspace{0.005\textwidth}
		\subfloat[Comparison of time consumption on IBM-Melbourne.]{\includegraphics[width=0.45\columnwidth]{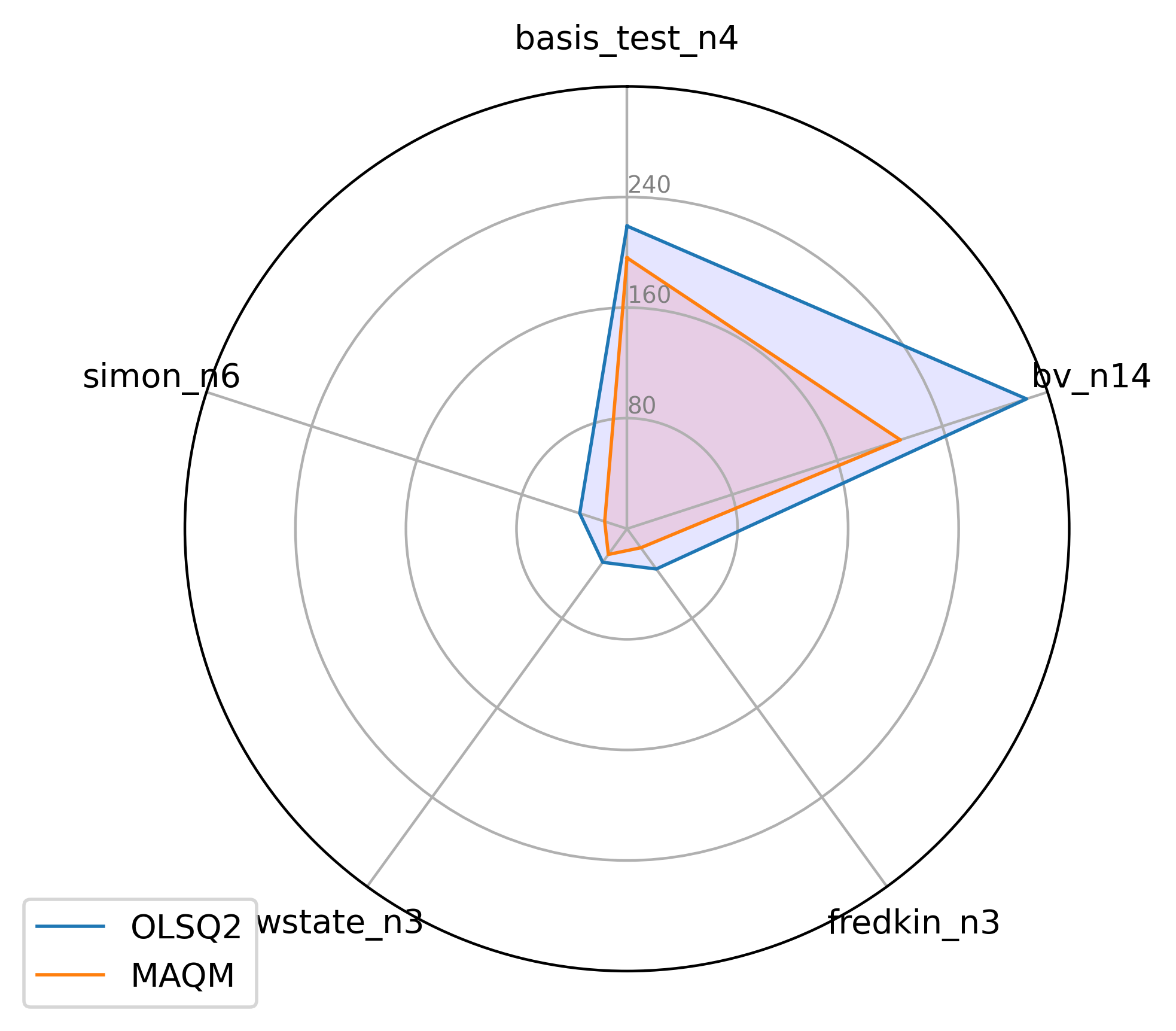} \label{ibmMelbourne_time}}
	\end{minipage}
	\\
	\centering
	\begin{minipage}[b]{\columnwidth}
		\subfloat[Comparison of time consumption on 5x5 grid.]{\includegraphics[width=0.45\columnwidth]{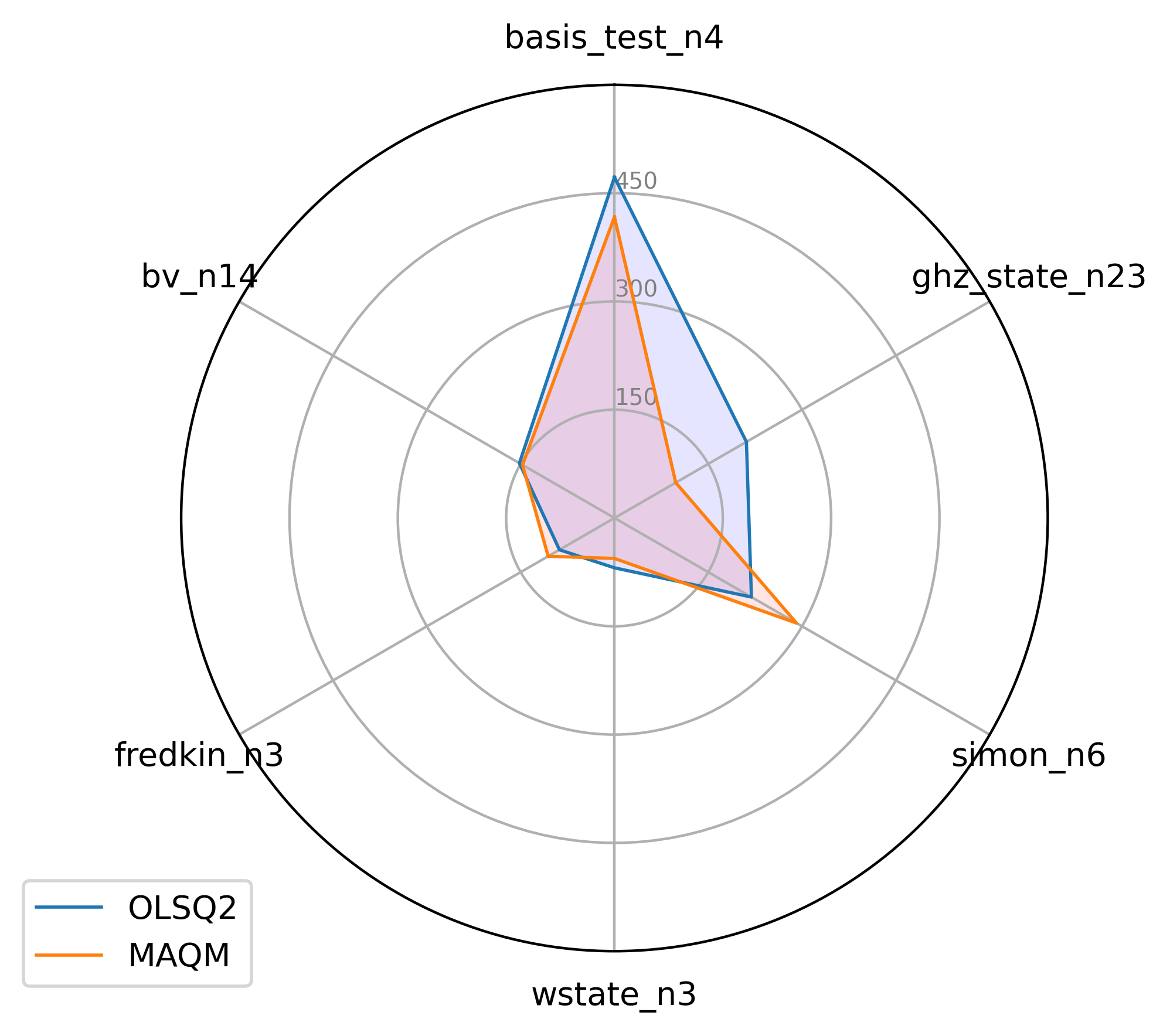} \label{5x5_time}} 
	\end{minipage}
	\caption{Comparison of time consumption.}
	\label{time}
\end{figure}

\subsection{MLQM versus OLSQ2}
To validate the efficiency of MLQM, we compare the runtime of MLQM with OLSQ2, the results of these samples are shown in Fig. \ref{time}. And the detailed time consumption for each sample is listed in the first column of Table \ref{tab:sycamore} to Table \ref{tab:5x5}, the numbers of searches are also shown in these tables. MLQM has an acceleration ratio of up to 6.78, which appears in the sample ``cat\_n35" on Sycamore. Generally, MLQM achieves an average acceleration ratio of 1.79, shows higher solution efficiency in most samples, especially for coupling graphs with more physical qubits and quantum circuit with more logical qubits, i.e., samples on Sycamore and IBM-Rochester. Furthermore, the number of searches for OLSQ2 increases dramatically with the number of qubits required for the sample on every coupling graphs, for example, quantum circuits with 14, 23, and 35 qubits require 17, 29 and 40 searches respectively processed by OLSQ2 method on Sycamore. 
In contrast, MLQM shows consistently excellent performance with a significant reduction in searches, even for those circuits and coupling graphs with many qubits, the average number of searches for MLQM is 6.56. By comparing the speed-up ratio of MLQM (displayed in parentheses after the time consumption of MLQM) with the percentage reduction in the number of searches of MLQM, it can be seen that samples with a higher percentage reduction of the number of searches also usually have higher speed-up ratios. The observation verifies our previous conjecture that it is reasonable and feasible to reduce the time consumption of qubit mapping by pruning the search space. Furthermore, in all cases, MLQM maintains no more than 12 searches, demonstrating excellent scalability. With respect to solution quality, both MLQM and OLSQ2 algorithms prioritize depth optimality in their initial search, subsequently refining solutions for swap number optimality, thereby ensuring MLQM's optimality.
\begin{samepage}
	\begin{table*}
		\small
		\caption{Comparison of OLSQ2 and MLQM on Sycamore.}
		\label{tab:sycamore}
		\centering
		\tabcolsep=0.015\linewidth
		\begin{tabular*}{\linewidth}{*{7}{c}}
			\toprule
			\multirow{2}*{Samples} &\multicolumn{2}{c}{\textbf{Time consumption (Seconds)}}  &\multicolumn{2}{c}{Memory (MB)} & \multicolumn{2}{c}{Search Counts}\\
			\cmidrule(lr){2-3}\cmidrule(lr){4-5}\cmidrule(lr){6-7}
			& OLSQ2 (DAC'23) & MLQM (Ours) & OLSQ2 (DAC'23) & MLQM (Ours) & OLSQ2 (DAC'23) & MLQM (Ours)\\
			\midrule
			wstate\_n3 &271 &\textbf{178} (1.52x)& 628 & \textbf{606} ($\downarrow$ 3.5\%) & 11 & \textbf{6}  ($\downarrow$ 45.5\%)\\
			bv\_n14 &605 &\textbf{458} (1.32x) & \textbf{1591} & 1605  ($\uparrow$ 0.9\%) & 17 & \textbf{8}  ($\downarrow$ 52.9\%)\\
			fredkin\_n3 &645 &\textbf{403} (1.6x) & 574 & \textbf{494}  ($\downarrow$ 14.1\%) & 9 & \textbf{8}  ($\downarrow$ 11.1\%)\\
			ghz\_state\_n23 &821 &\textbf{180} (4.56x) & 1568 & \textbf{1157}  ($\downarrow$ 26.2\%) & 29 & \textbf{4}  ($\downarrow$ 86.2\%)\\
			simon\_n6 &\textbf{1119} &1580 (0.71x) & \textbf{1447} & 1469  ($\uparrow$ 1.5\%) & 11 & \textbf{7}  ($\downarrow$ 36.3\%)\\
			basis\_test\_n4 &1199 &\textbf{885} (1.35x) & 5369 & \textbf{3286} ($\downarrow$ 38.8\%)) & 6 & \textbf{4}  ($\downarrow$ 33.3\%)\\
			cat\_n35 &5774 &\textbf{851} (6.78x) & 4018 & \textbf{2293}  ($\downarrow$ 42.9\%) & 40 & \textbf{4}  ($\downarrow$ 90.0\%)\\
			ghz\_n40 &15282 &\textbf{2706} (5.65x) & 4615 & \textbf{3131}  ($\downarrow$ 32.2\%) & 44 & \textbf{6}  ($\downarrow$ 86.4\%)\\
			\bottomrule
		\end{tabular*}
	\end{table*}
	\begin{table*}
		\small
		\caption{Comparison of OLSQ2 and MLQM on IBM-Rochester.}
		\label{tab:IBMRochester}
		\centering
		\tabcolsep=0.015\linewidth
		\begin{tabular*}{\linewidth}{*{7}{c}}
			\toprule
			\multirow{2}*{Samples} &\multicolumn{2}{c}{\textbf{Time consumption (Seconds)}}  &\multicolumn{2}{c}{Memory (MB)} & \multicolumn{2}{c}{Search Counts}\\
			\cmidrule(lr){2-3}\cmidrule(lr){4-5}\cmidrule(lr){6-7}
			& OLSQ2 (DAC'23) & MLQM (Ours) & OLSQ2 (DAC'23) & MLQM (Ours) & OLSQ2 (DAC'23) & MLQM (Ours)\\
			\midrule
			bv\_n14 &44326 &\textbf{32904} (1.34x) & 4269 & \textbf{3267}  ($\downarrow$ 23.5\%) & 21 & \textbf{12} ($\downarrow$ 42.9\%)\\
			cat\_n35 &4689 &\textbf{2398} (1.96x) & 2952 & \textbf{2340}  ($\downarrow$ 20.7\%) & 43 & \textbf{8}  ($\downarrow$ 81.4\%)\\
			ghz\_n40 &7747 &\textbf{1313} (5.9x) & 4036 & \textbf{2420}  ($\downarrow$ 40\%) & 37 & \textbf{5}  ($\downarrow$ 86.5\%)\\
			wstate\_n3 &194 &\textbf{109} (1.08x)& \textbf{438} & 454 ($\uparrow$ 3.7\%) & \textbf{7} & \textbf{7} (0\%)\\
			fredkin\_n3 &132 &\textbf{78} (1.69x) & 411 & \textbf{343}  ($\downarrow$ 16.5\%) & 8 & \textbf{6} ($\downarrow$ 25\%)\\
			ghz\_state\_n23 &474 &\textbf{418} (1.13x) & 1159 & \textbf{978}  ($\downarrow$ 15.6\%) & 24 & \textbf{7}  ($\downarrow$ 70.83\%)\\
			basis\_test\_n4 &690 &\textbf{634} (1.09x) & 2731 & \textbf{2636} ($\downarrow$ 3.5\%) & \textbf{3} & 5  ($\uparrow$ 66.7\%)\\
			simon\_n6 &\textbf{552} &1336 (0.41x) & \textbf{805} & 837  ($\downarrow$ 4\%) & 11 & \textbf{8}  ($\downarrow$ 27.3\%)\\
			\bottomrule
		\end{tabular*}
	\end{table*}
	\begin{table*}
		\small
		\caption{Comparison of OLSQ2 and MLQM on Aspen-4.}
		\label{tab:aspen4}
		\centering
		\tabcolsep=0.015\linewidth
		\begin{tabular*}{\linewidth}{*{7}{c}}
			\toprule
			\multirow{2}*{Samples} &\multicolumn{2}{c}{\textbf{Time consumption (Seconds)}}  &\multicolumn{2}{c}{Memory (MB)} & \multicolumn{2}{c}{Search Counts}\\
			\cmidrule(lr){2-3}\cmidrule(lr){4-5}\cmidrule(lr){6-7}
			& OLSQ2 (DAC'23) & MLQM (Ours) & OLSQ2 (DAC'23) & MLQM (Ours) & OLSQ2 (DAC'23) & MLQM (Ours)\\
			\midrule
			bv\_n14 & 188& \textbf{134} (1.4x) & 669 & \textbf{451} ($\downarrow$ 32.5\%) & 11 & \textbf{8}  ($\downarrow$ 27.3\%)\\
			basis\_test\_n4 & 204& \textbf{181} (1.13x)& 1029 & \textbf{839} ($\downarrow$ 18.5\%) & \textbf{3} & 4  ($\uparrow$ 33.3\%)\\
			wstate\_n3 & 30& \textbf{22} (1.36x)& \textbf{209} & 213 ($\uparrow$ 1.6\%) & 9 & \textbf{6}  ($\downarrow$ 33.3\%)\\
			fredkin\_n3 & 30& \textbf{16} (1.84x)& \textbf{188} & 207  ($\uparrow$ 9.8\%) & 8 & \textbf{6}  ($\downarrow$ 25.0\%)\\
			simon\_n6 & 64& \textbf{58} (1.1x)& \textbf{362} & 375  ($\uparrow$ 3.9\%) & \textbf{7} & 8  ($\uparrow$ 14.3\%)\\
			\bottomrule
		\end{tabular*}
	\end{table*}
	\begin{table*}
		\small
		\caption{Comparison of OLSQ2 and MLQM on IBM-Melbourne.}
		\label{tab:IBMMelbourne}
		\centering
		\tabcolsep=0.015\linewidth
		\begin{tabular*}{\linewidth}{*{7}{c}}
			\toprule
			\multirow{2}*{Samples} &\multicolumn{2}{c}{\textbf{Time consumption (Seconds)}}  &\multicolumn{2}{c}{Memory (MB)} & \multicolumn{2}{c}{Search Counts}\\
			\cmidrule(lr){2-3}\cmidrule(lr){4-5}\cmidrule(lr){6-7}
			& OLSQ2 (DAC'23) & MLQM (Ours) & OLSQ2 (DAC'23) & MLQM (Ours) & OLSQ2 (DAC'23) & MLQM (Ours)\\
			\midrule
			bv\_n14 & 304 & \textbf{208} (1.46x)& 707 & \textbf{449} ($\downarrow$ 36.5\%) & 13 & \textbf{8}  ($\downarrow$ 38.5\%)\\
			simon\_n6 & 70& \textbf{57} (1.23x)& 381 & \textbf{374}  ($\downarrow$ 1.8\%) & \textbf{7} & \textbf{7}  (0\%)\\
			basis\_test\_n4 & 219& \textbf{196} (1.12x)& 1095 & \textbf{855} ($\downarrow$ 21.9\%) & \textbf{3} & 6  ($\uparrow$ 100.0\%)\\
			wstate\_n3 & 30 & \textbf{23} (1.3x) & \textbf{211} & 224 ($\uparrow$ 6.4\%) & 9 & \textbf{6}  ($\downarrow$ 33.3\%)\\
			fredkin\_n3 & 36& \textbf{17} (2.2x)& \textbf{198} & 205  ($\uparrow$ 3.6\%) & 8 & \textbf{6}  ($\downarrow$ 25.0\%)\\
			\bottomrule
		\end{tabular*}
	\end{table*}
	\begin{table*}
		\small
		\caption{Comparison of OLSQ2 and MLQM on 5x5 coupling architecture.}
		\label{tab:5x5}
		\centering
		\tabcolsep=0.015\linewidth
		\begin{tabular*}{\linewidth}{*{7}{c}}
			\toprule
			\multirow{2}*{Samples} &\multicolumn{2}{c}{\textbf{Time consumption (Seconds)}}  &\multicolumn{2}{c}{Memory (MB)} & \multicolumn{2}{c}{Search Counts}\\
			\cmidrule(lr){2-3}\cmidrule(lr){4-5}\cmidrule(lr){6-7}
			& OLSQ2 (DAC'23) & MLQM (Ours) & OLSQ2 (DAC'23) & MLQM (Ours) & OLSQ2 (DAC'23) & MLQM (Ours)\\
			\midrule
			basis\_test\_n4 & 472& \textbf{417} (1.13x)& 1982 & \textbf{1545} ($\downarrow$ 22\%) & \textbf{3} & 4  ($\uparrow$ 33.3\%)\\
			bv\_n14 & 152 & \textbf{147} (1.03x)& \textbf{656} & 840 ($\uparrow$ 28\%) & \textbf{8} & \textbf{8}  (0\%)\\
			wstate\_n3 & 69 & \textbf{56} (1.23x) & \textbf{316} & 355 ($\uparrow$ 12.3\%) & 9 & \textbf{6}  ($\downarrow$ 33.3\%)\\
			ghz\_state\_n23 & 211 & \textbf{98} (2.2x) & \textbf{653} & 681 ($\uparrow$ 4.3\%) & 17 & \textbf{6}  ($\downarrow$ 64.7\%)\\
			fredkin\_n3 & \textbf{88}& 106 (0.83x)& \textbf{281} & 285  ($\uparrow$ 1.4\%) & \textbf{8} & \textbf{8}  (0\%)\\
			simon\_n6 & \textbf{219} & 291 (0.75x)& 688 & \textbf{654}  ($\downarrow$ 4.9\%) & \textbf{7} & 8  ($\uparrow$ 14.3\%)\\
			\bottomrule
		\end{tabular*}
	\end{table*}
\end{samepage}

Table \ref{tab:sycamore} to Table \ref{tab:5x5} also present the memory footprint comparison between MLQM and OLSQ2, highlighting MLQM's space efficiency. With an average 22\% reduction in memory usage, MLQM demonstrates a smaller memory footprint than OLSQ2 in most samples. 
The memory reduction is especially prominent on coupling graphs with more physical qubits. On Sycamore, OLSQ2 consumed up to 5369 MB of memory, while MLQM required only 3286 MB, marking a significant 38.8\% reduction. MLQM also exhibits remarkable space efficiency across various architectures, achieving substantial memory footprint reductions for samples exceeding 4,000 MB on OLSQ2. Similar to the solving efficiency, the spatial efficiency improvement of the MLQM also exhibits an increase with logic and physical qubits, which underscores MLQM's robust scalability concerning space efficiency, particularly as the memory requirements increase. 

\subsection{MLQM versus SABRE}
\begin{figure}[tbp]
	\vspace{0.8em}
	\centering
	\includegraphics[width=\columnwidth]{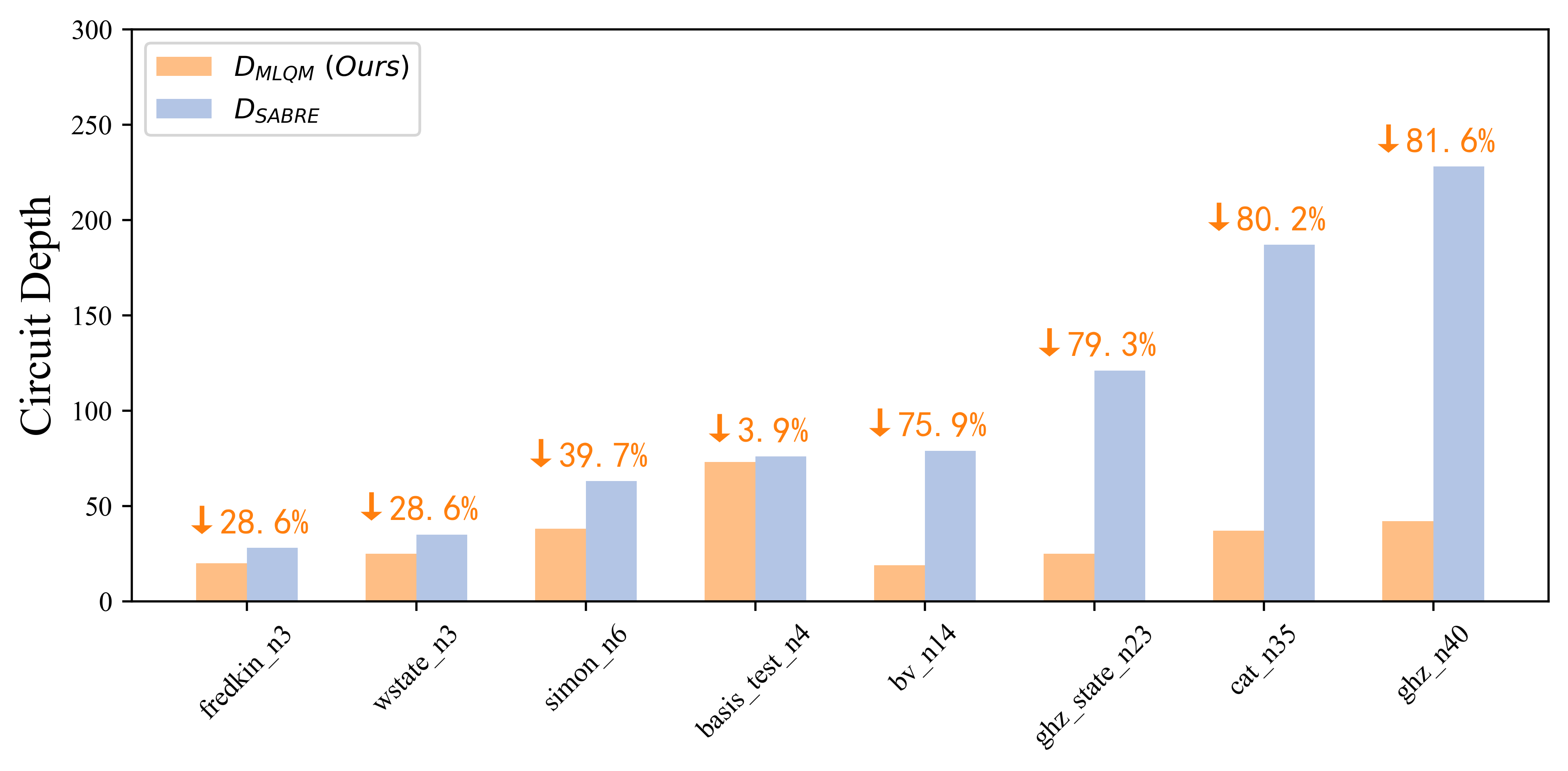}
	\vspace{-2.5em}
	\caption{Comparison of the circuit depth after qubit mapping between MLQM and SABRE on Sycamore, the depth reduction percentage are shown at the top of bars.}
	\label{fig:syca_depth}
\end{figure}

\begin{figure}[tbp]
	\centering 
	\includegraphics[width=\columnwidth]{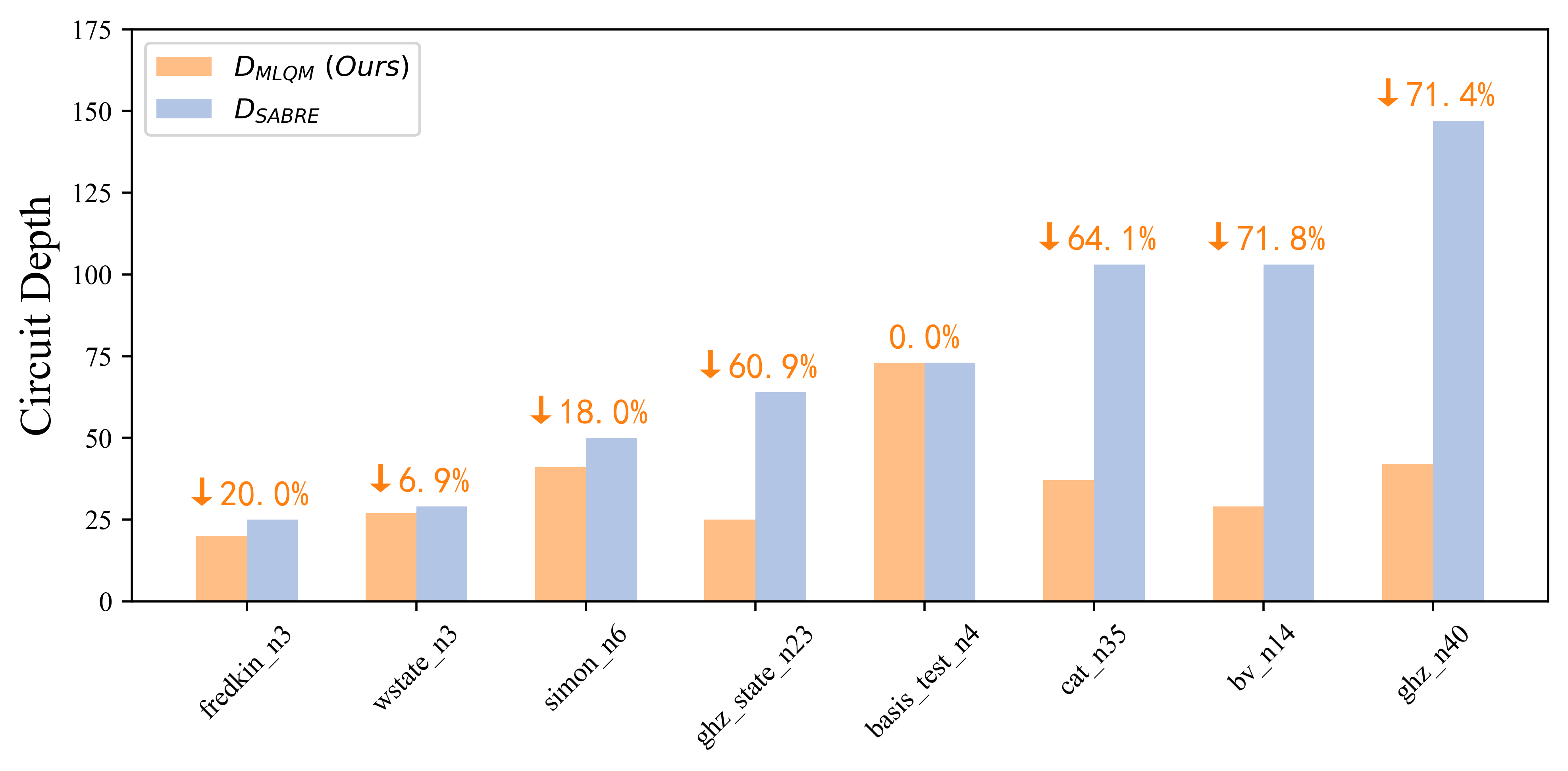}
	\vspace{-2.5em}
	\caption{Comparison of the circuit depth after qubit mapping between MLQM and SABRE on IBM-Rochester, the depth reduction percentage are shown at the top of bars.}
	\label{fig:ibmRochester_depth}
\end{figure}

\begin{figure}[htbp]
	\centering
	\includegraphics[width=\columnwidth]{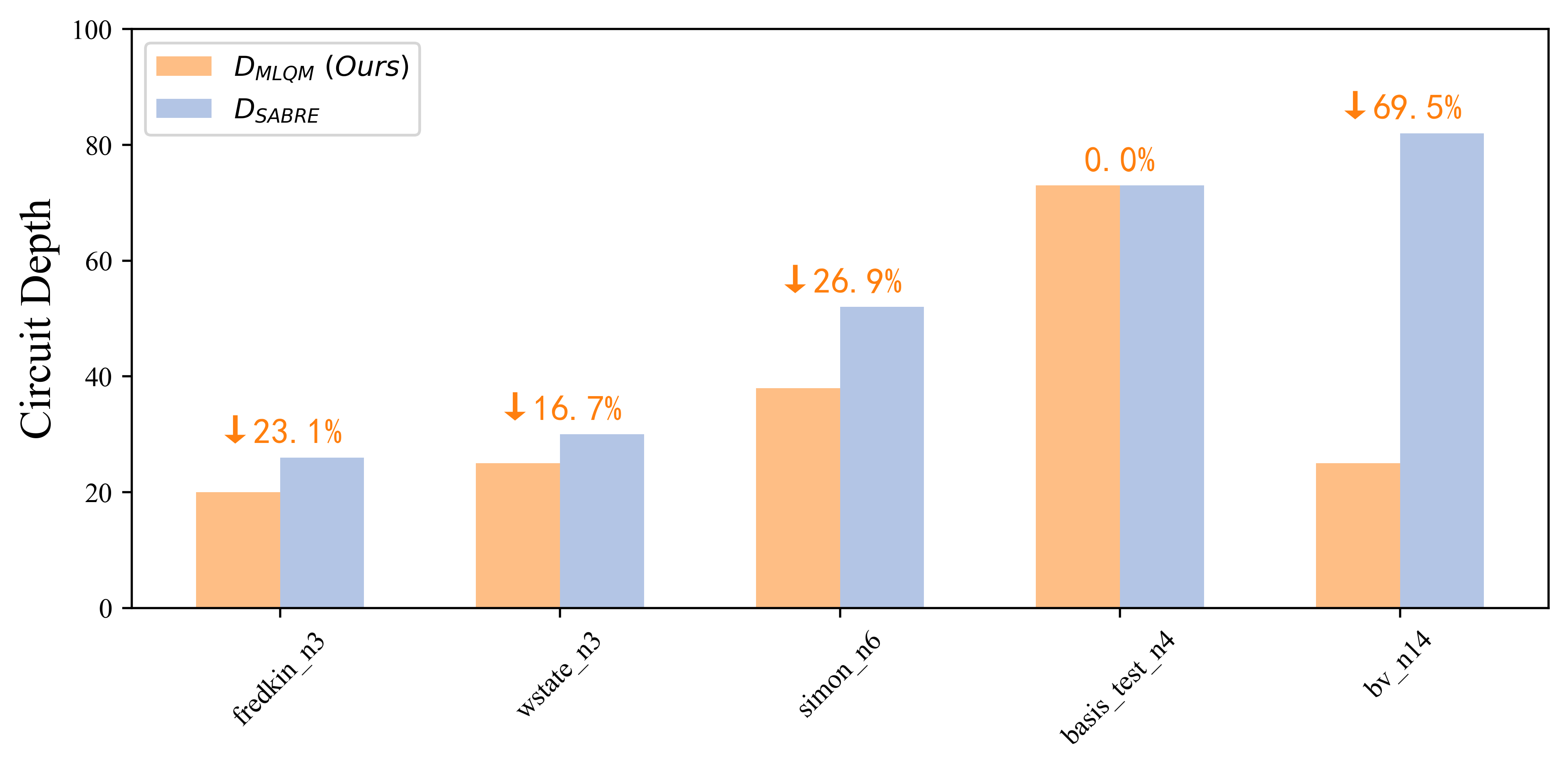}
	\vspace{-2.5em}
	\caption{Comparison of the circuit depth after qubit mapping between MLQM and SABRE on Aspen-4, the depth reduction percentages are shown at the top of bars.}
	\label{fig:aspen4_depth}
\end{figure}

\begin{figure}[htbp]
	\centering
	\includegraphics[width=\columnwidth]{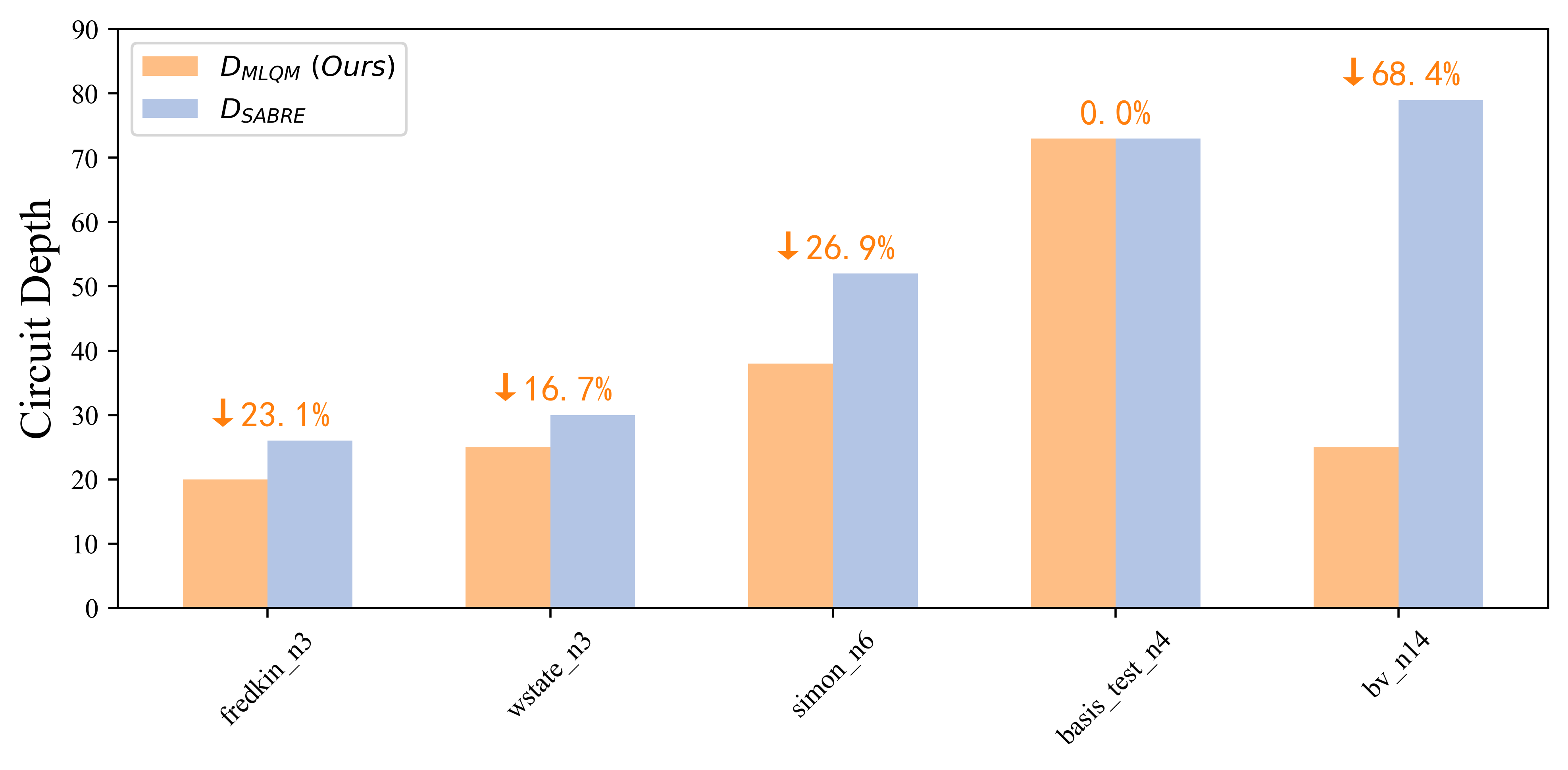}
	\vspace{-2.5em}
	\caption{Comparison of the circuit depth after qubit mapping between MLQM and SABRE on IBM-Melbourne, the depth reduction percentages of MLQM relative to SABRE are shown at the top of bars.}
	\label{fig:ibmMelbourne_depth}
\end{figure}

\begin{figure}[bp]
	\centering
	\includegraphics[width=\columnwidth]{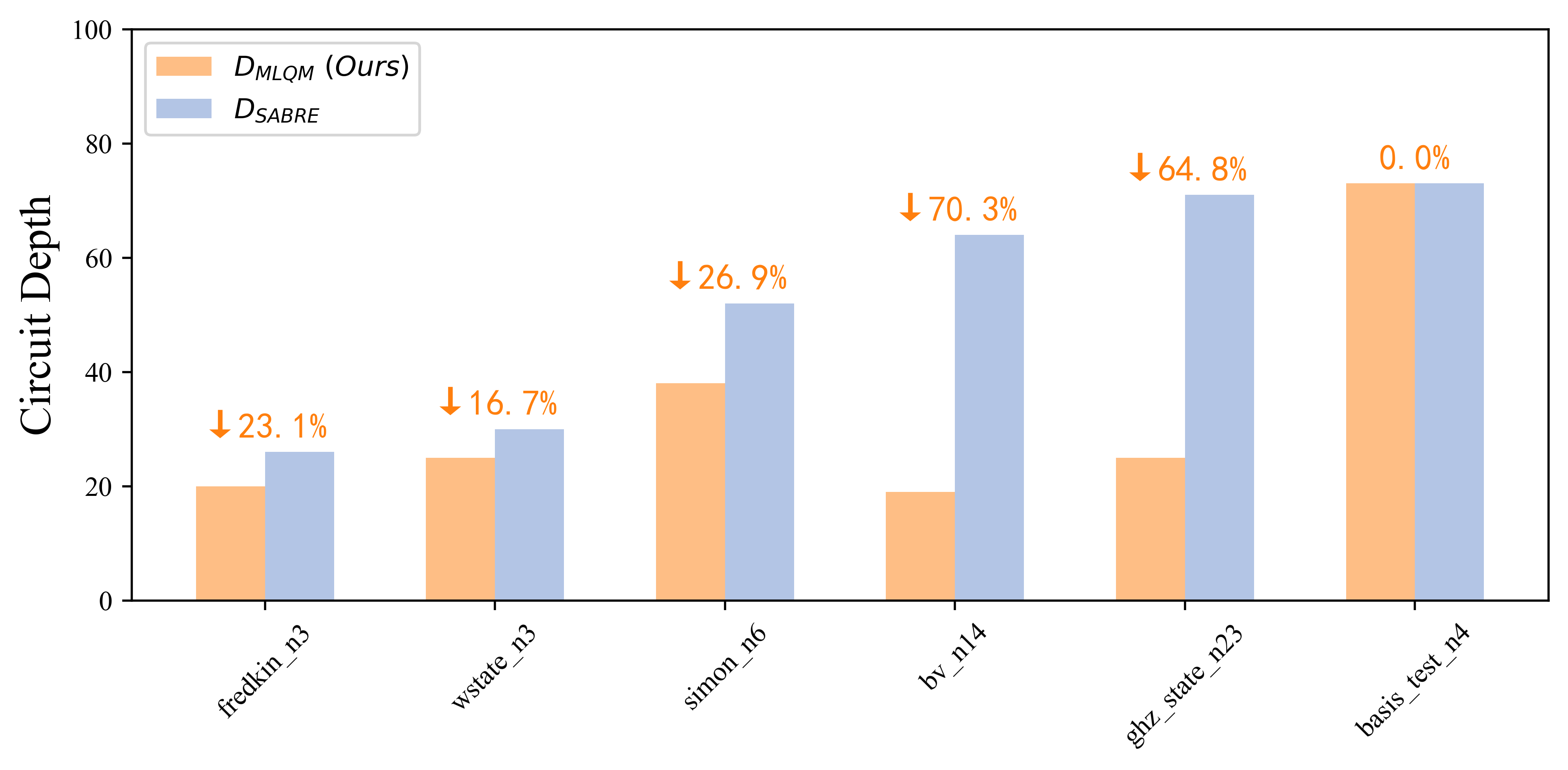}
	\vspace{-2.5em}
	\caption{Comparison of the circuit depth after qubit mapping between MLQM and SABRE on the 5x5 coupling architecture, the depth reduction percentages of MLQM relative to SABRE are shown at the top of bars.}
	\label{fig:5x5_depth}
\end{figure}

To demonstrate the quality of MLQM, the mapped circuit depths of MLQM with the heuristic method SABRE\cite{liTacklingQubitMapping2019} are compared, and the results are shown in Fig. \ref{fig:syca_depth} to Fig. \ref{fig:5x5_depth}. MLQM achieves an average of 52.2\%, 39.1\%, 27.2\%, 27.0\% and 33.6\% depth reductions on Sycamore, IBM-Rochester, Aspen-4, IBM-Melbourne and 5x5 grid architecture respectively, and the overall average depth reduction across all coupling graphs is 35.8\%. It is worth noting that as the coupling graph's qubit number increases, the depth improvement of MLQM becomes more and more obvious. Similarly, the advantage of high-quality solutions provided by MLQM is more pronounced as the number of qubits of the quantum circuits increases for a fixed coupling architecture, which is clearly demonstrated in most figures.

\begin{figure}[tbp]
	\centering 
	\includegraphics[width=\columnwidth]{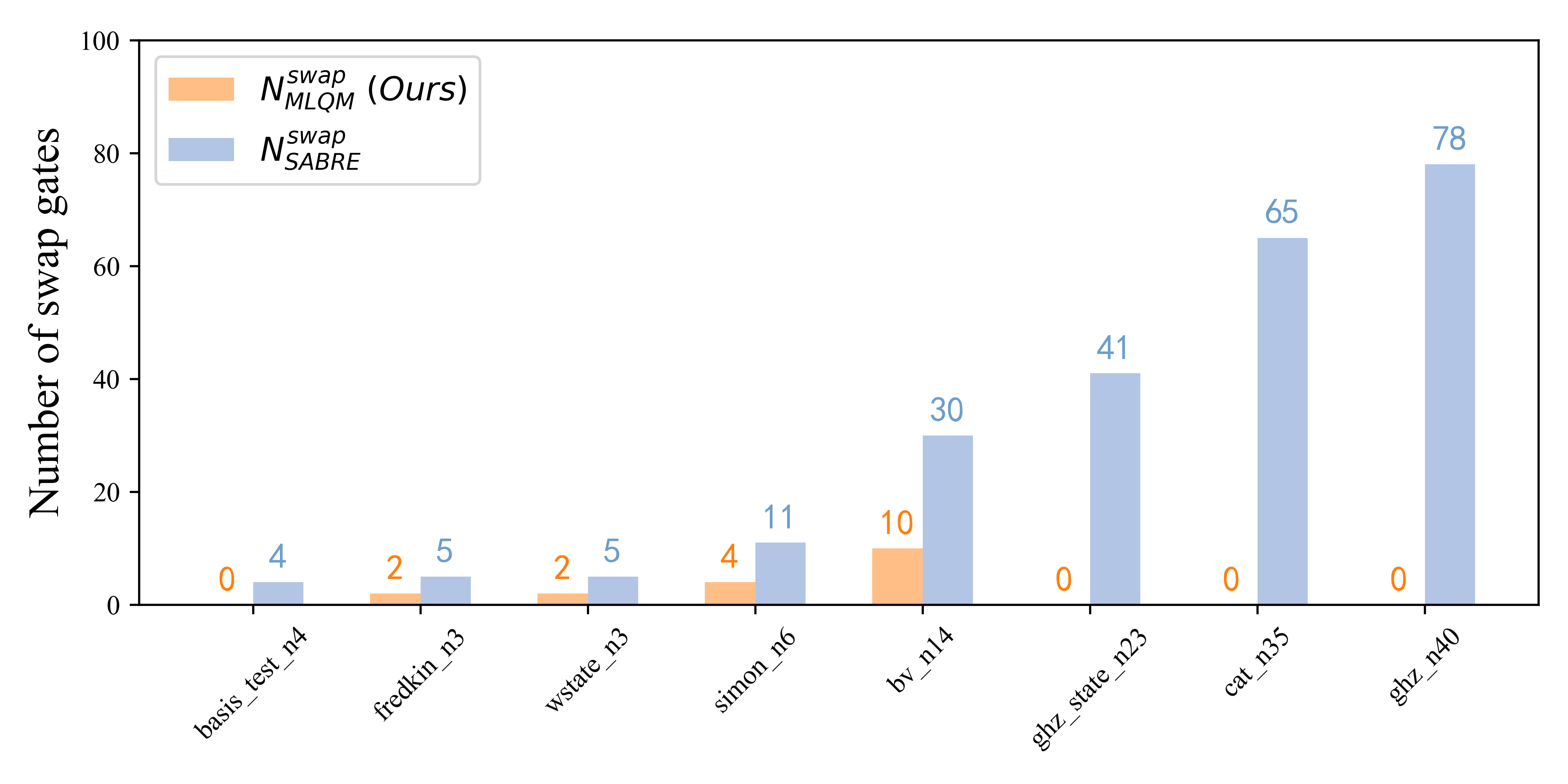}
	\vspace{-2.5em}
	\caption{Comparison of the swap gate number after qubit mapping between MLQM and SABRE on Sycamore.}
	\label{fig:syca_swapnum}
\end{figure}

\begin{figure}[tbp]
	\centering
	\includegraphics[width=\columnwidth]{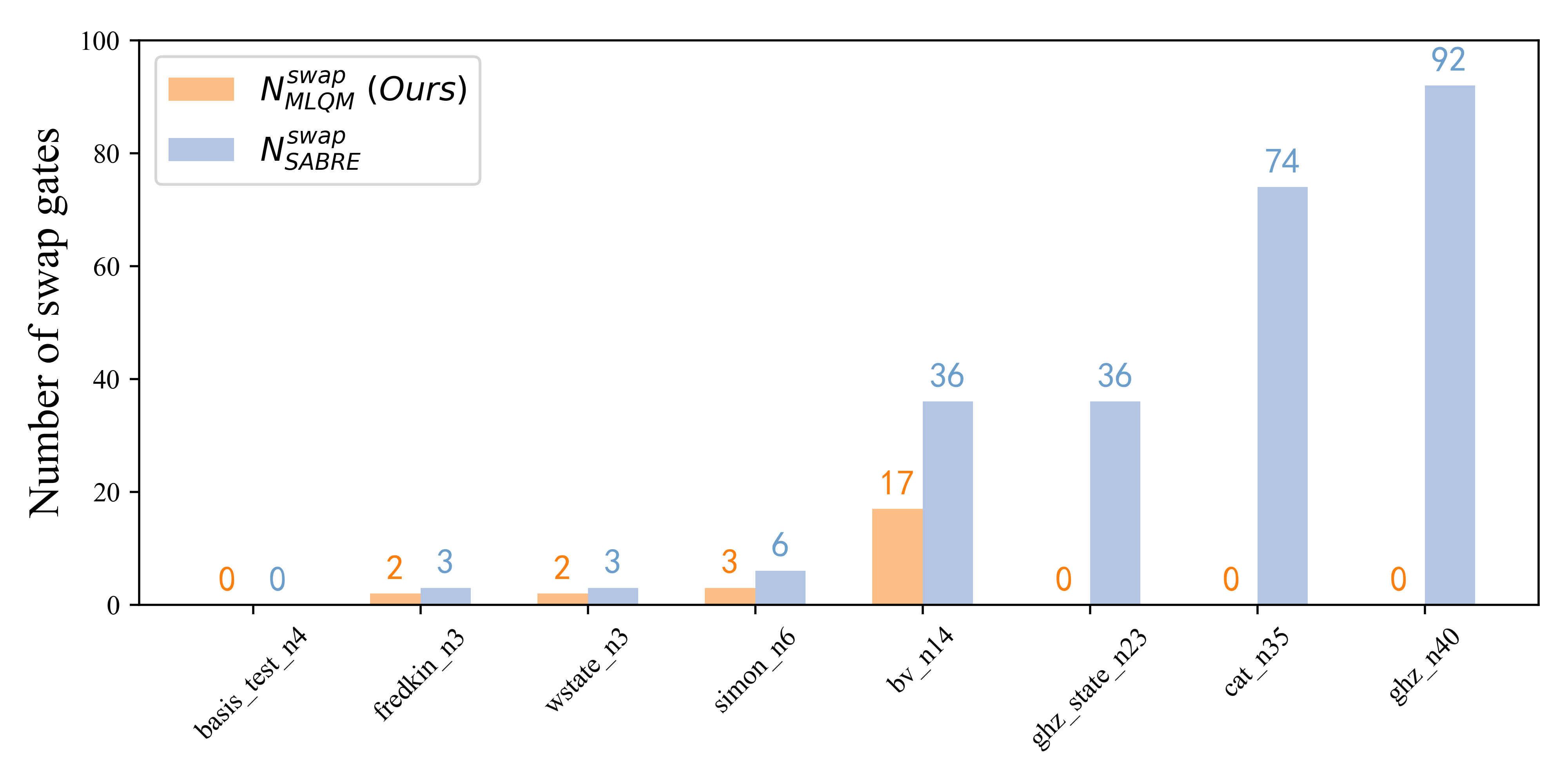}
	\vspace{-2.5em}
	\caption{Comparison of the swap gate number after qubit mapping between MLQM and SABRE on IBM-Rochester.}
	\label{fig:ibmRochester_swapnum}
\end{figure}

\begin{figure}[tbp]
	\centering
	\includegraphics[width=\columnwidth]{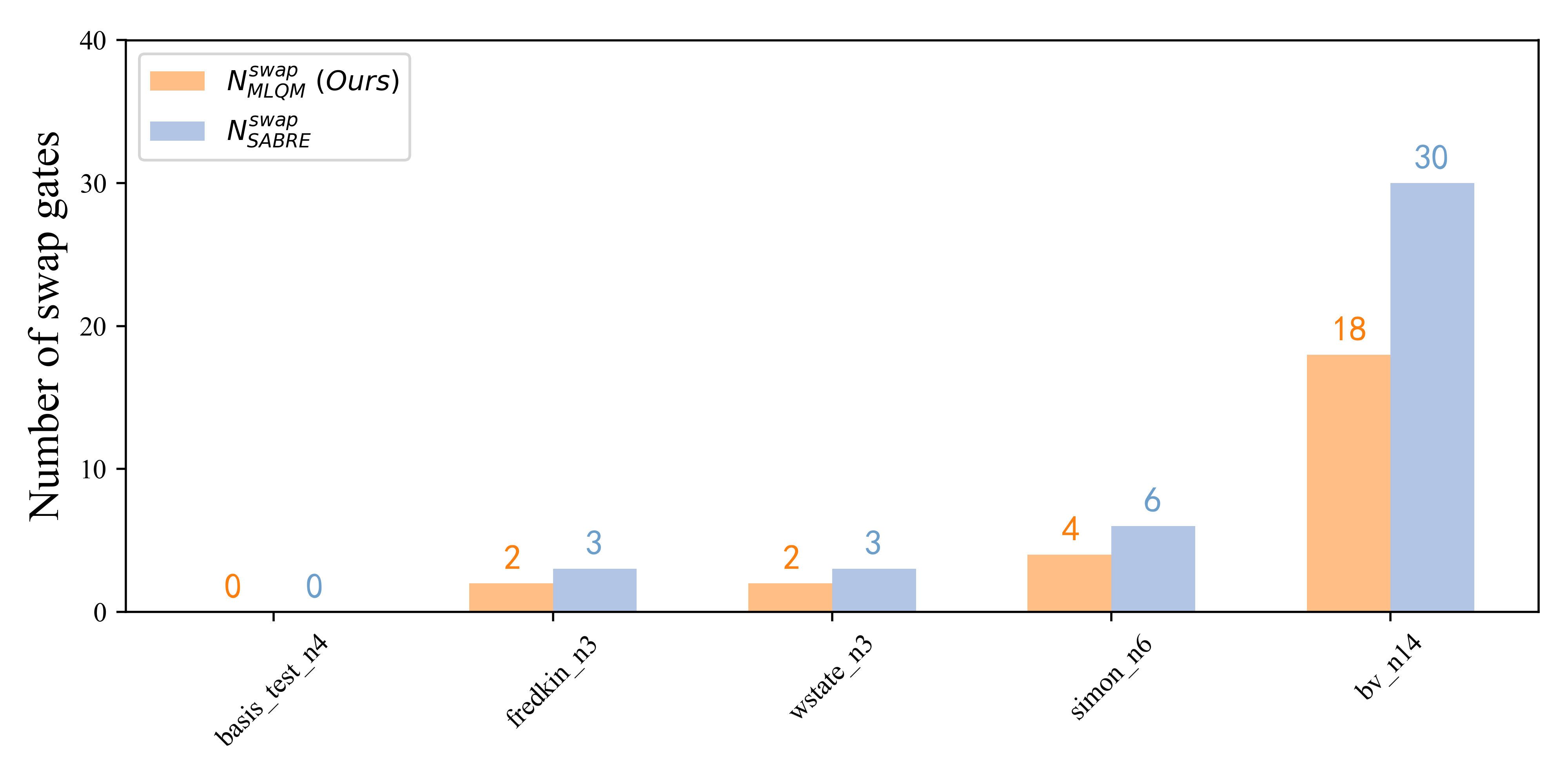}
	\vspace{-2.5em}
	\caption{Comparison of the swap gate number after qubit mapping between MLQM and SABRE on Aspen-4.}
	\label{fig:aspen4_swapnum}
\end{figure}

\begin{figure}[tbp]
	\centering
	\includegraphics[width=\columnwidth]{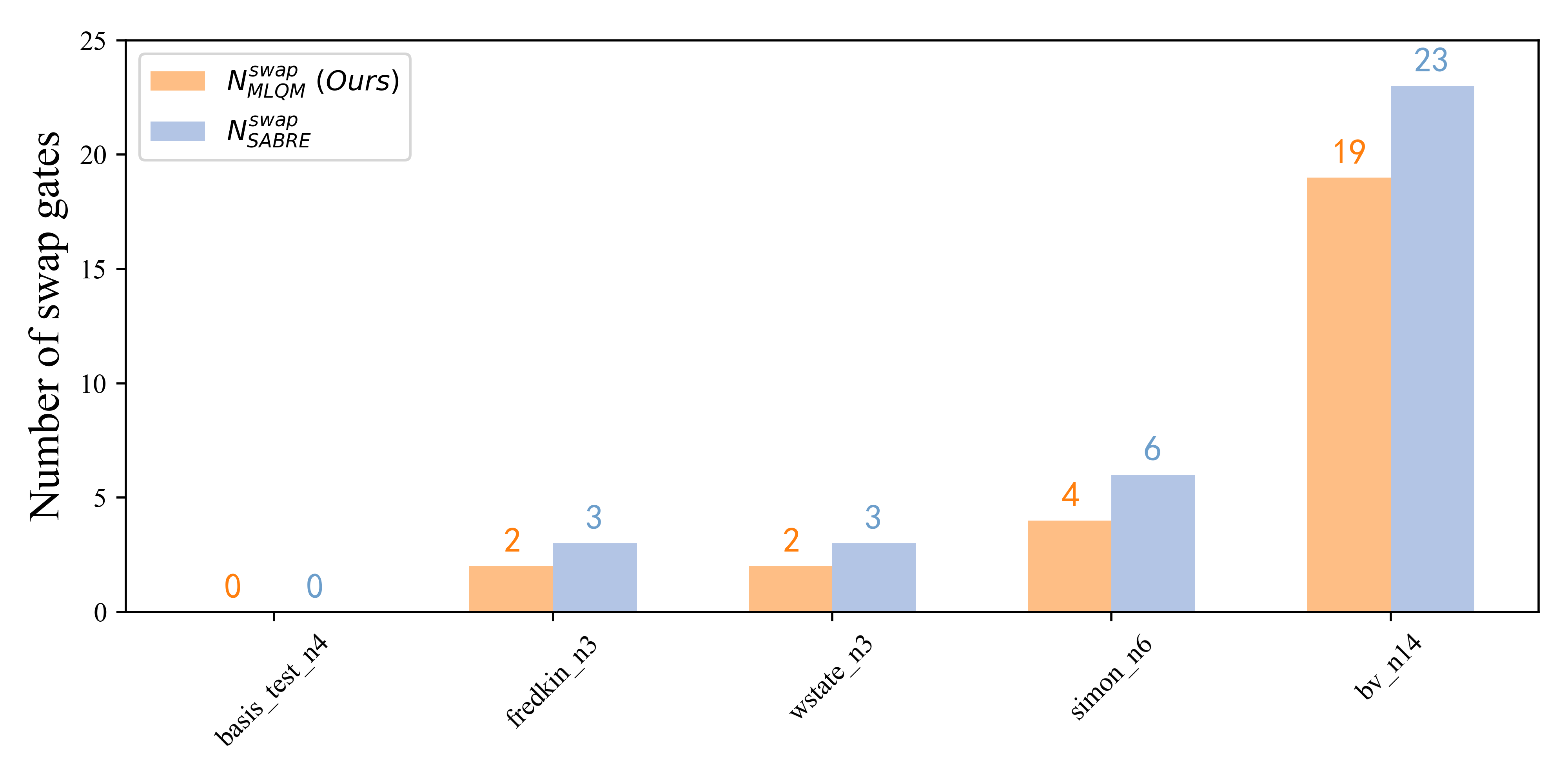}
	\vspace{-2.5em}
	\caption{Comparison of the swap gate number after qubit mapping between MLQM and SABRE on IBM-Melbourne.}
	\label{fig:ibmMelbourne_swapnum}
\end{figure}

\begin{figure}[tbp]
	\centering
	\vspace{1.1em}
	\includegraphics[width=\columnwidth]{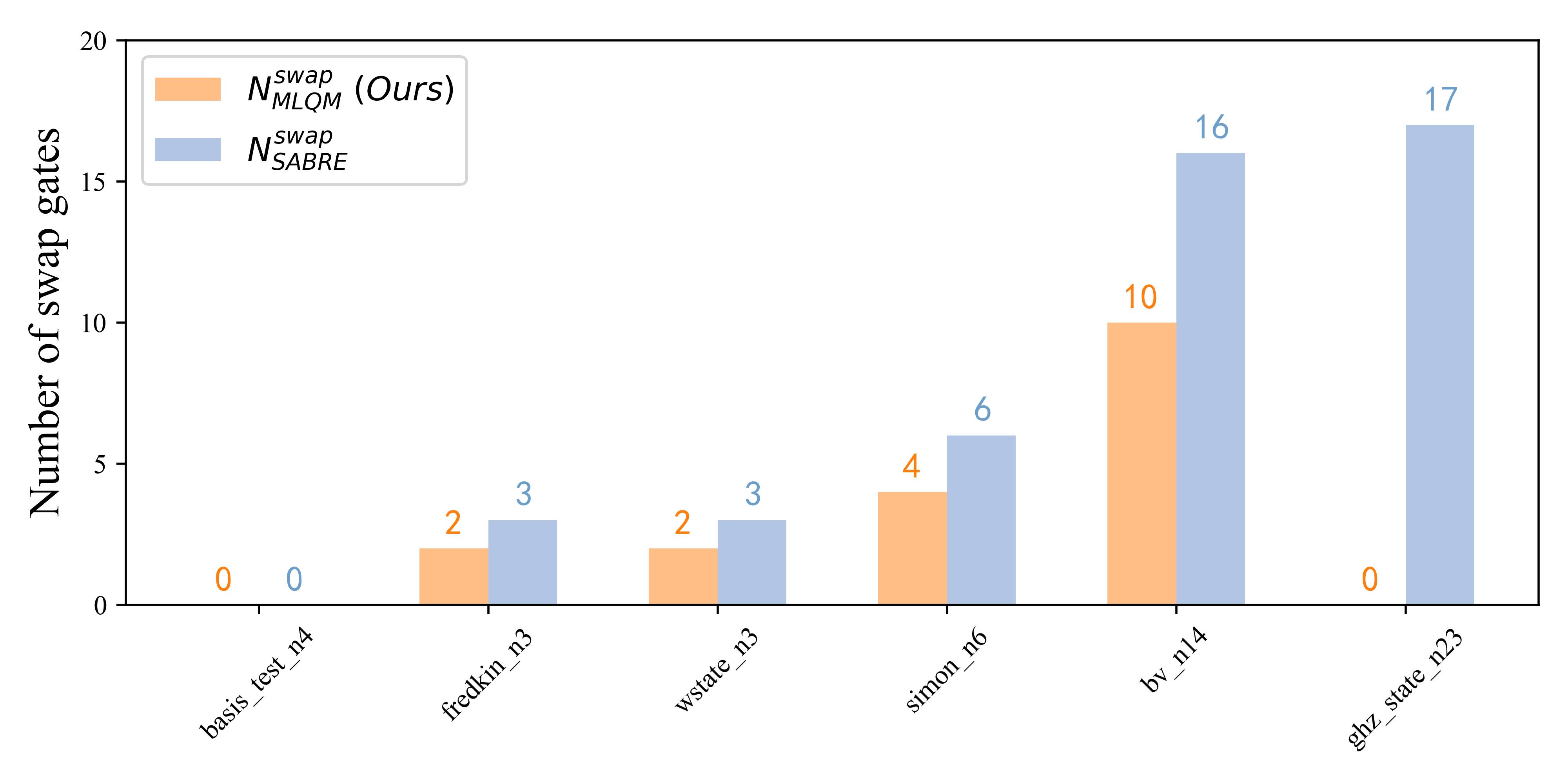}
	\vspace{-2.5em}
	\caption{Comparison of the swap gate number after qubit mapping between MLQM and SABRE on 5x5 coupling architecture.}
	\label{fig:5x5_swapnum}
\end{figure}

\begin{figure*}[htbp]
	\centering
	\subfloat[Result of SABRE.]{\includegraphics[width=0.92\textwidth]{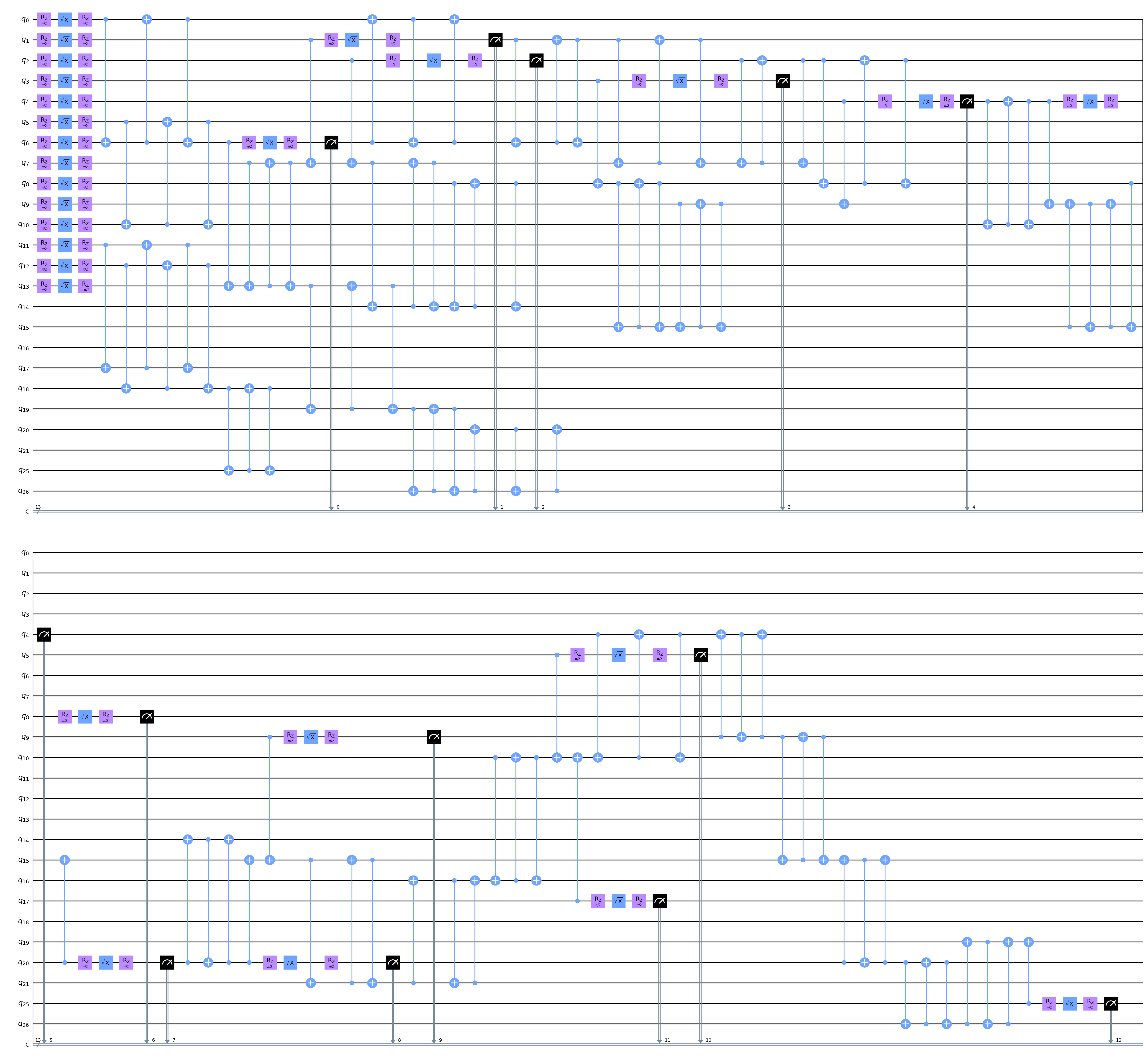}\label{sabre_circ}}
	\\
	\subfloat[Result of MLQM.]{\includegraphics[width=0.92\textwidth]{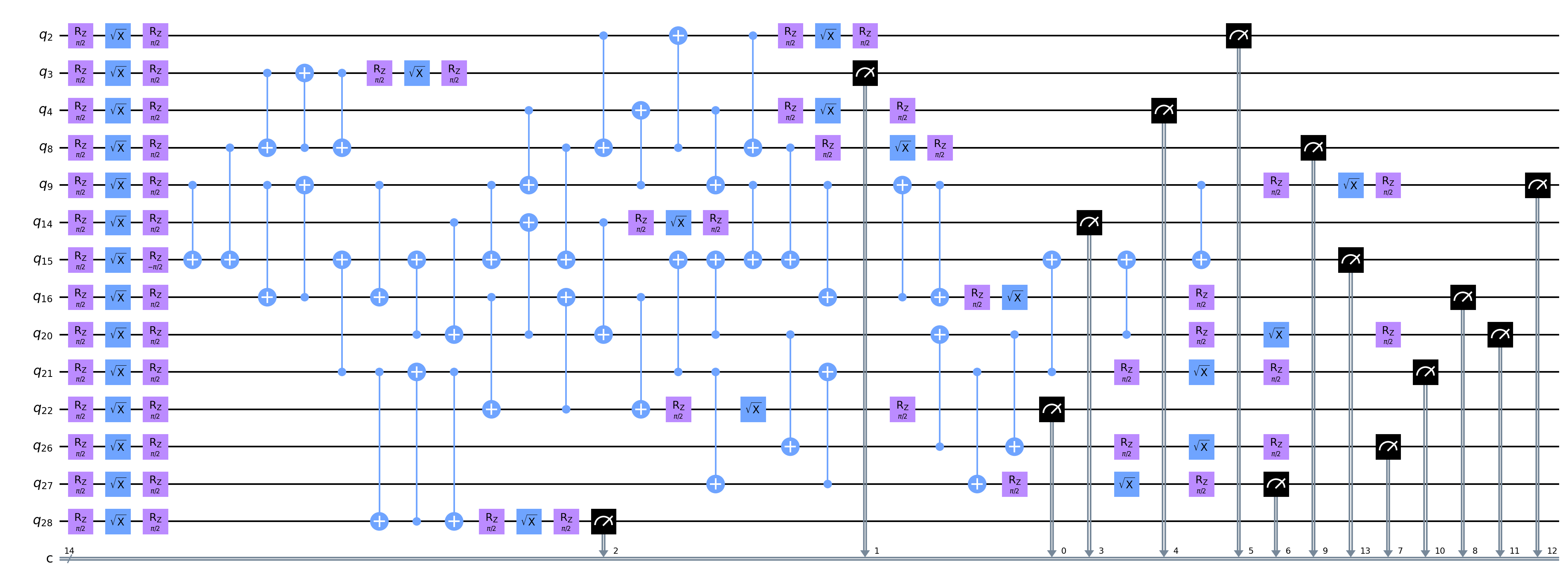}\label{mlqm_circ}}
	\caption{Intuitive comparison of MLQM and SABRE on Sycamore.}
	\label{two_circuits}
\end{figure*}

The number of swap gates used in qubit mapping with SABRE is compared to the proposed MLQM approach across various architectures in Fig. \ref{fig:syca_swapnum} to Fig. \ref{fig:5x5_swapnum}. MLQM achieves a significant reduction in the average number of swap gates by 81.3\%, 58.7\%, 28.0\%, 23.5\%, and 39.6\% on the Sycamore, IBM-Rochester, Aspen-4, IBM-Melbourne, and 5x5 grid architectures, respectively. The overall average depth reduction across all coupling graphs is 46.2\%. The result shows that the competitiveness of the solutions of MLQM is increasing for coupling graphs with more physical qubits. 
Notably, MLQM attains a 100\% reduction in the number of swap gates for seven samples on the Sycamore, IBM-Rochester, and 5x5 grid architectures, indicating that no additional gates are introduced, thus preserving circuit fidelity. 
To demonstrates the effectiveness of MLQM more intuitively, a comparison between SABRE and MLQM on the ``bv\_n14" are shown in sample (Fig. \ref{two_circuits}\subref{sabre_circ} and Fig. \ref{two_circuits}\subref{mlqm_circ}). Comparing with SABRE, MLQM produces more compact circuits, higher qubit utilization, and significantly reduced circuit lengths. In addition, it also avoids the extensive use of auxiliary qubits in SABRE. In general, MLQM transforms quantum programs into shorter and more compact circuits that use fewer qubits. The advancement contributes to enhancing the success rate of quantum circuit on real quantum devices, reducing the resource utilization demands on quantum computers, and augmenting the parallel processing capabilities of quantum programs.


\section{Discussion}
MLQM proposes a new solution to address the trade-off between accuracy and efficiency encountered by conventional qubit mapping methods, contributing to the resolution of the prevalent scalability issue that optimal qubit mapping algorithms face. This work provides novel research perspectives and benchmarks for further optimization of quantum circuit mapping algorithms, demonstrating the potential for application on larger-scale quantum circuits and more complex quantum computer architectures, thus facilitating the practicality of quantum computing. 
Based on this work, future research could explore machine learning techniques to adjust constraints, simplifying the problem space and providing insights to aid reasoning, particularly for problems near the satisfiability boundary. Moreover, as quantum hardware advances, integrating machine learning, fidelity data, and solver-based methods shows great potential for developing more efficient qubit mapping techniques with higher scalability and fidelity.

\section{Conclusion}
This work proposes a machine learning-based approach for accelerating optimal qubit mapping utilizing quantum circuit feature datasets. By leveraging an augmented and refined quantum circuit dataset, MLQM employs machine learning techniques to provide high-quality prior knowledge to the solver, significantly reducing the solver's global search space, thereby accelerating the solving process. Moreover, MLQM adaptively adjusts the constraints in the solver, further narrowing the local search space. Consequently, MLQM eliminates the reliance on manually setting solver constraints based on experience, greatly enhancing the efficiency of qubit mapping.

Experimental results conducted on five quantum computer hardware architectures demonstrate the robust efficiency and stability of MLQM. Compared to OLSQ2, the state-of-the-art solver-based method, MLQM achieves an average speedup ratio of 1.79, with a maximum speed-up of 6.78. In terms of memory footprint, MLQM achieves an average reduction of 22\%.
Compared with the leading heuristic algorithm SABRE, MLQM reduces the circuit depth by an average of 35.8\% and the number of swap gates by an average of 46.2\%, furthermore, the advantage of MLQM increases with the number of qubits in quantum circuits and coupling graphs, which demonstrates the excellent quality of the solutions provided by MLQM.

\section*{Acknowledgements}
This work was supported in part by the National Natural Science Foundation of China under Grant 62472072, the National Natural Science Foundation of China under Grant 62172075, and in part by Natural Science Foundation of Xinjiang Uygur Autonomous Region of China under Grant 2023D01A63.

\clearpage
\clearpage

\appendix
\section{Examples and Analysis of Regression Trees}
\label{app:regression tree}
\begin{figure}[htbp]
	\centering
	\subfloat[Tree predicting circuit depth.]{\includegraphics[width=0.2\textwidth]{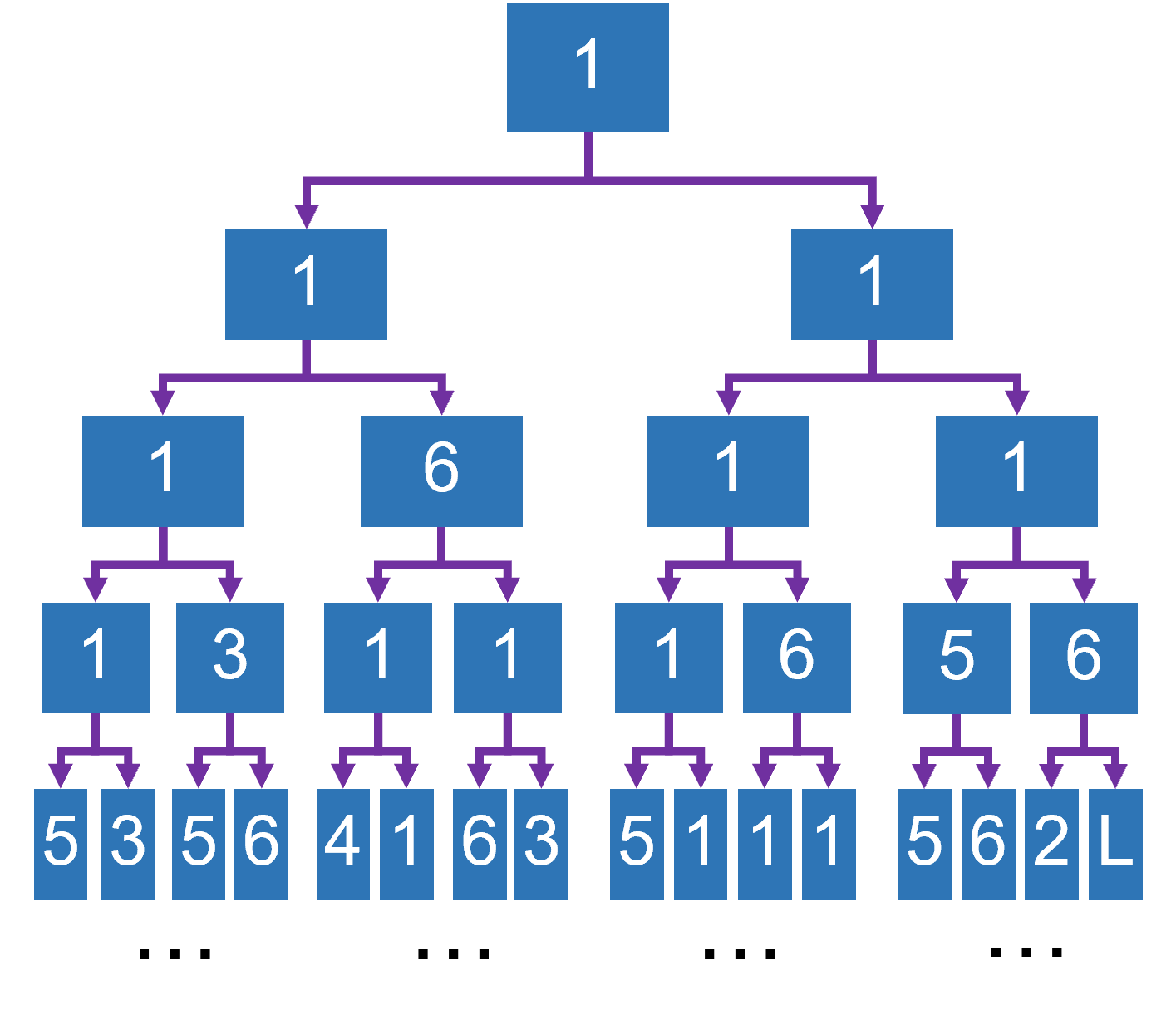}\label{fig:tree_ibmrochester_depth}}
	\hspace{0.04\textwidth}
	\subfloat[Tree predicting swap number.]{\includegraphics[width=0.2\textwidth]{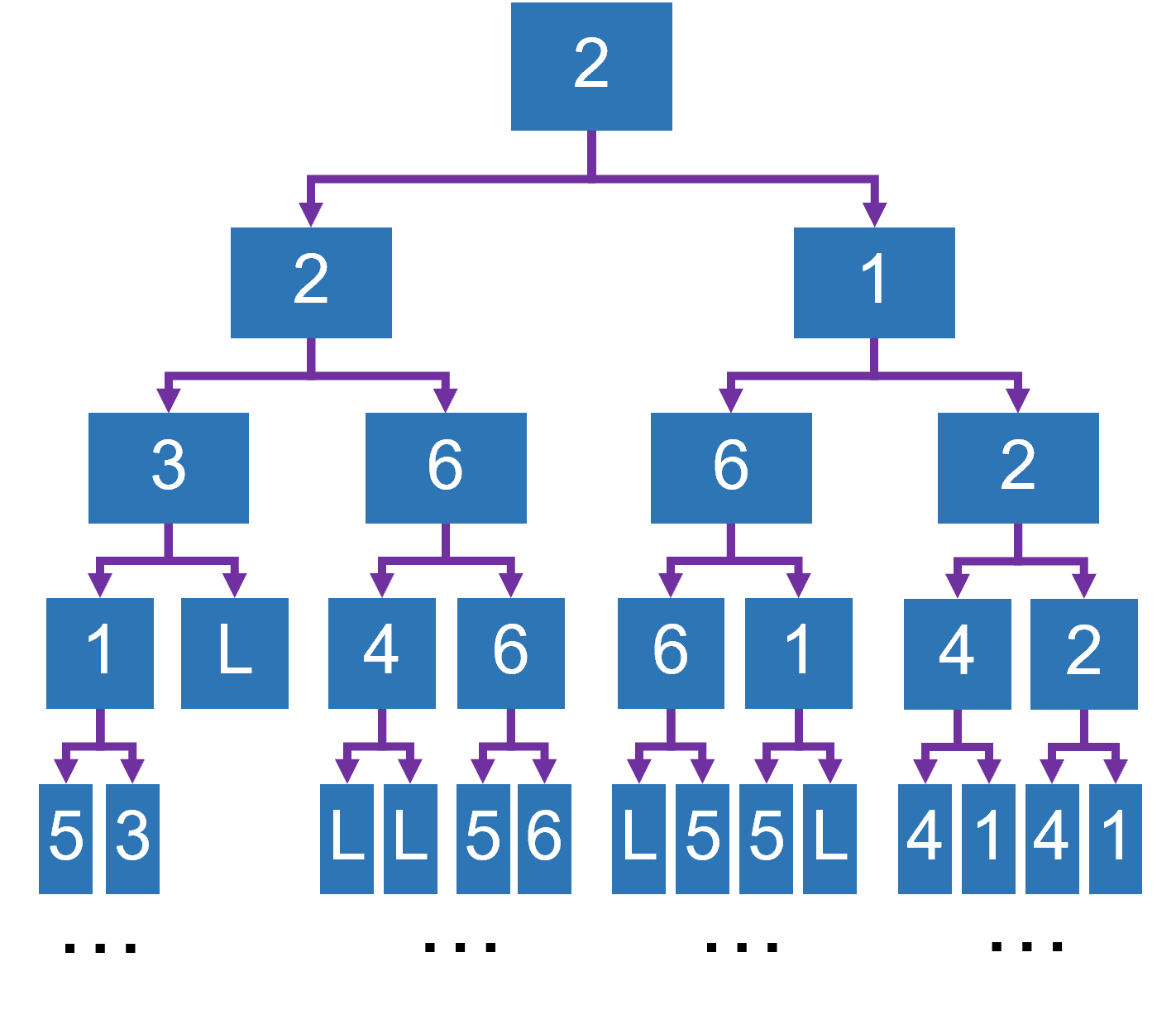}\label{fig:tree_ibmrochester_swapnum}}
	\caption{Structure of regression trees for IBM Rochester. Nodes represent split features: 1 for circuit depth, 2 for circuit width, 3 for max qubit depth, 4 for operation density, 5 for two-qubit gate count, 6 for entanglement variance, and L for leaf nodes.}
	\label{fig:tree_ibmrochester}
\end{figure}

\begin{figure}[htbp]
	\centering
	\subfloat[Tree predicting circuit depth.]{\includegraphics[width=0.2\textwidth]{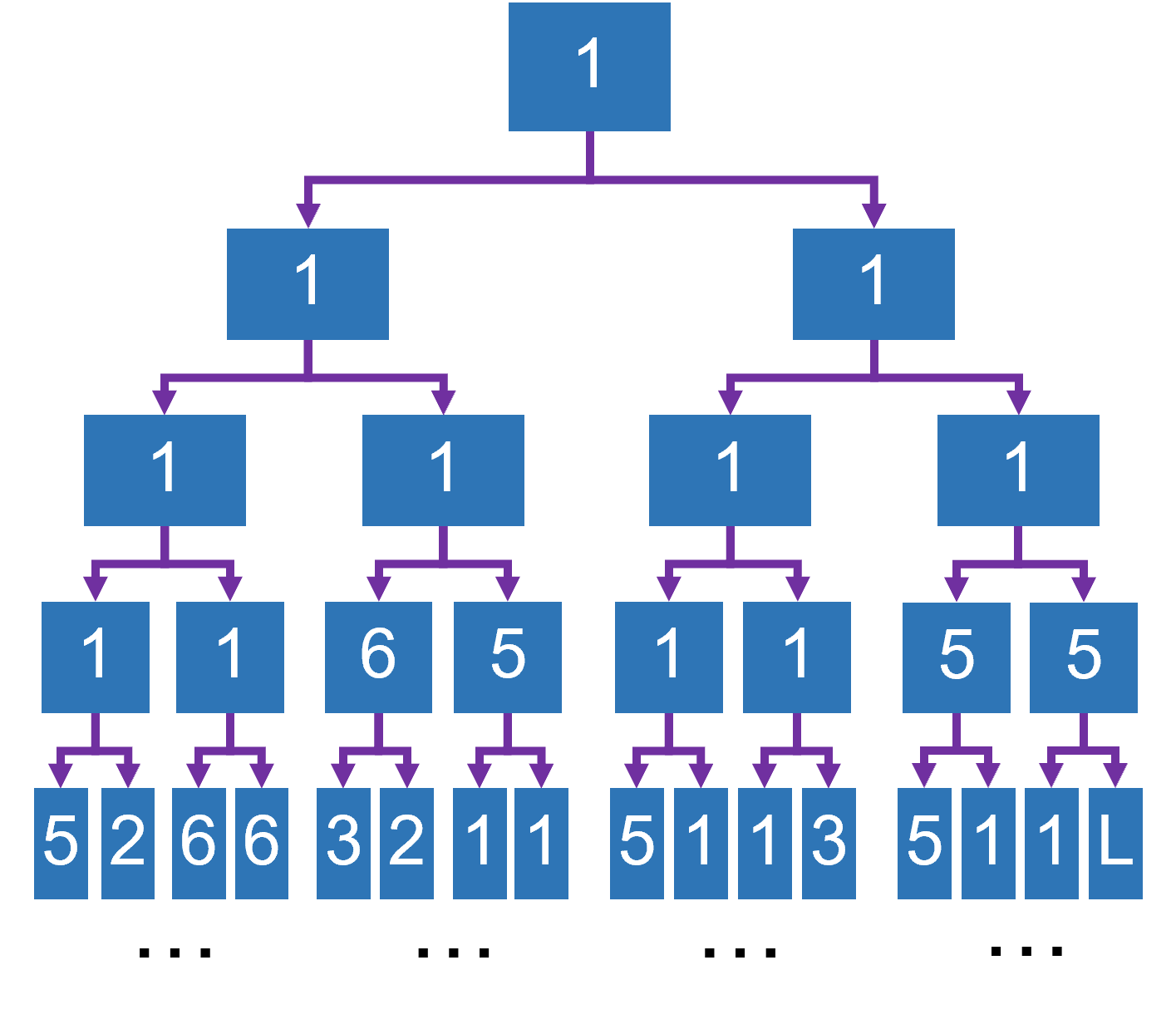}\label{fig:tree_ibmmelbourne_depth}}
	\hspace{0.04\textwidth}
	\subfloat[Tree predicting swap number.]{\includegraphics[width=0.2\textwidth]{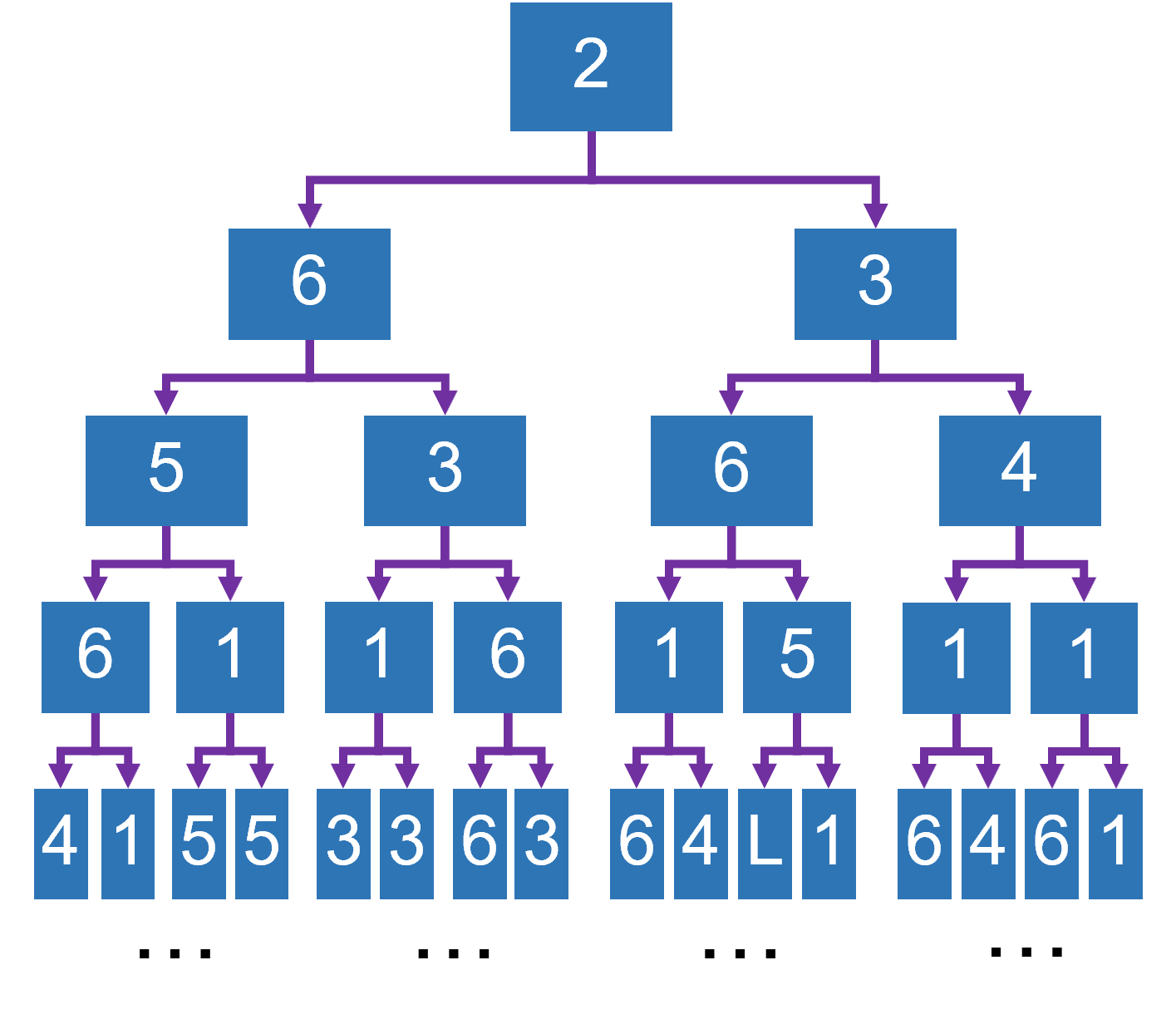}\label{fig:tree_ibmmelbourne_swapnum}}
	\caption{Structure of regression trees for IBM Melbourne: Nodes represent split features: 1 for circuit depth, 2 for circuit width, 3 for max qubit depth, 4 for operation density, 5 for two-qubit gate count, 6 for entanglement variance, and L for leaf nodes.}
	\label{fig:tree_melbourne}
\end{figure}

\begin{figure}[tbp]       
	\centering
	\includegraphics[width=0.98\columnwidth]{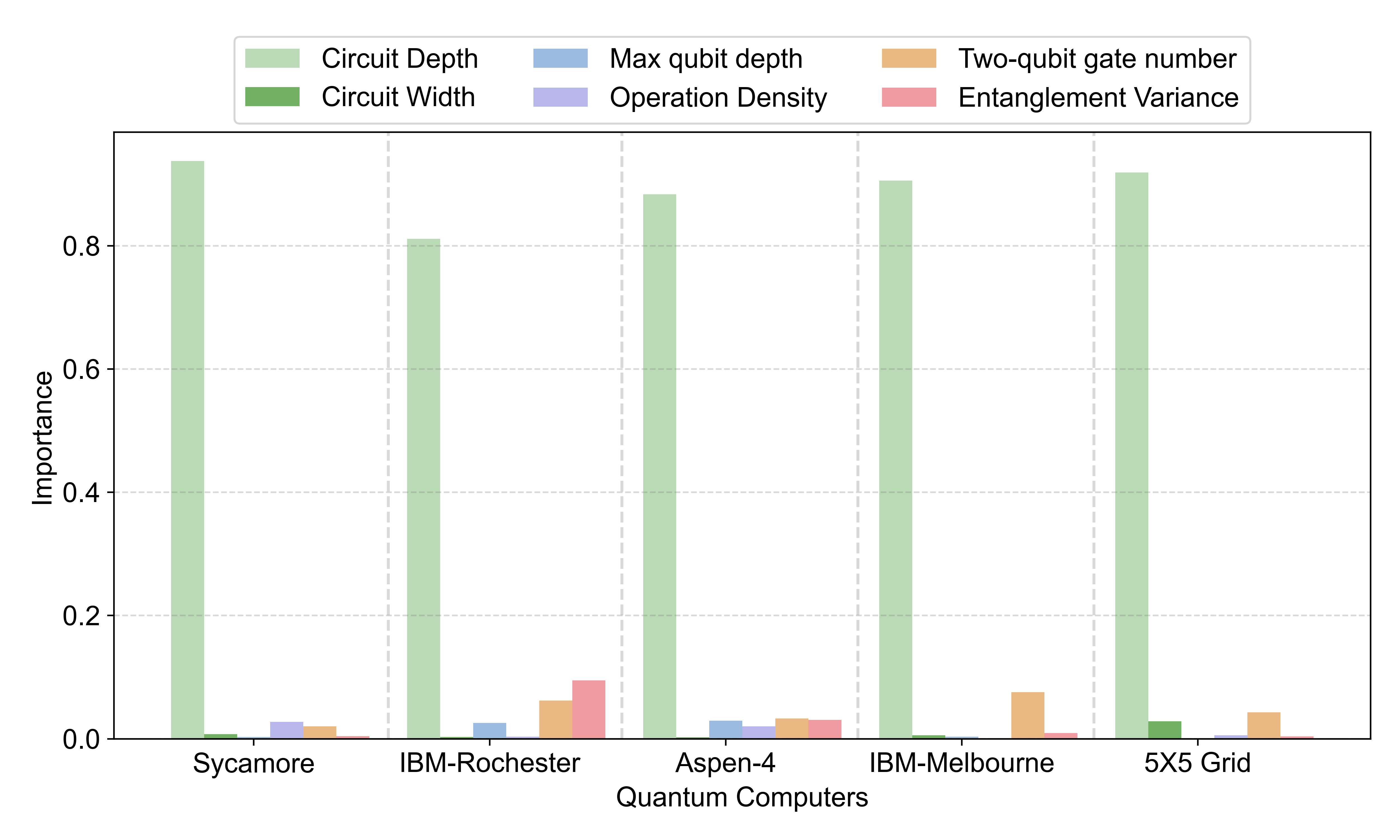}
	\caption{Feature importance in predicting circuit depth after qubit mapping.}
	\label{fig:importance_depth}
\end{figure}

\begin{figure}[tbp]
	\centering
	\includegraphics[width=0.98\columnwidth]{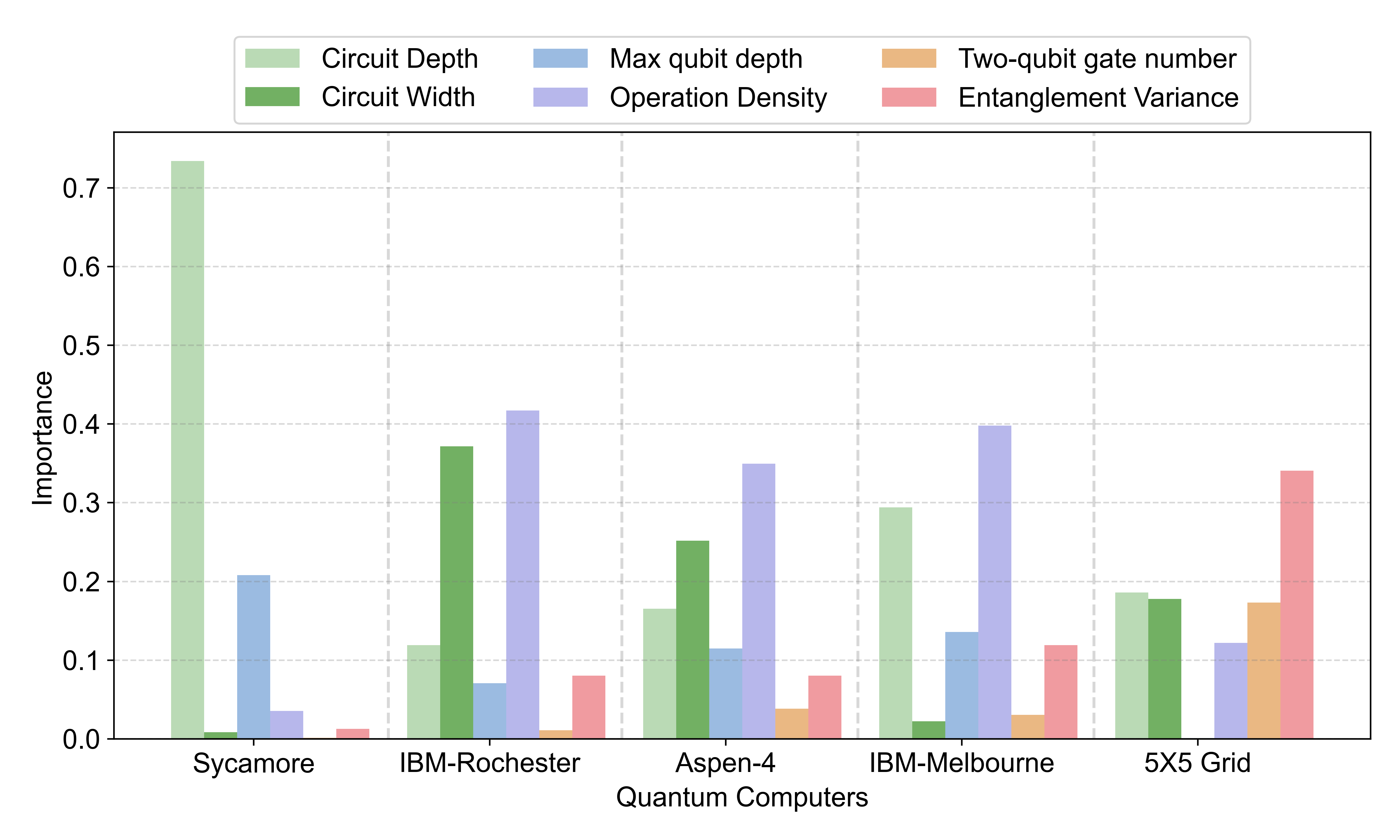}
	\caption{Feature importance in predicting swap number after qubit mapping.}
	\label{fig:importance_swapnum}
\end{figure}

This appendix presents two sets of results. The first set displays the tree structures of the models trained on circuit mapping for IBM Rochester and IBM Melbourne, as shown in Figures \ref{fig:tree_ibmrochester} and \ref{fig:tree_melbourne}, respectively. The second set illustrates the feature importance for the depth model and swap number model across all quantum computers, as depicted in Figures \ref{fig:importance_depth} and \ref{fig:importance_swapnum}. The feature importance is computed using the following formulas:

\begin{equation}
	\label{importance}
	Importance(f_{j})=\frac{W(f_{j})}{\sum{}{} W(f)}.
\end{equation}

\begin{equation}
	\label{importance_w}
	W(f_{j})=\sum_{n_{i}\in N_{f_{j}}}{}(\text{MSE}_{n_{i}}-(\frac{n_{i,l}}{n_{i}} \cdot \text{MSE}_{n_{i,l}}+\frac{n_{i,r}}{n_{i}} \cdot \text{MSE}_{n_{i,r}})).
\end{equation}
Here, $f_{j}$ denotes the target feature, $N_{f_{j}}$ are nodes where $f_{j}$ is used as the splitting feature, $\text{MSE}_{n_{i}}$ denotes the mean squared error at node $n_{i}$, and $n_{i,l}$ and $n_{i,r}$ are the left and right children of node $n_{i}$. These data provide intriguing insights and lead to several significant conclusions.


Fig. \ref{fig:importance_depth} shows circuit depth is the most important feature on all quantum computer when predicting circuit depth after qubit mapping. When the coupling graph is a fully connected graph, qubit mapping does not introduce any additional swap gates, so the circuit depth after mapping remains the same as the original circuit depth. As the connectivity of the coupling graph decreases, the number of swap gates required by the circuit gradually increases, causing the mapped circuit depth to deviate from the original circuit depth. Sycamore and the 5x5 grid have the strongest connectivity, which explains why circuit depth is most critical for them.


According to Fig. \ref{fig:importance_swapnum}, circuit depth is the most important feature for Sycamore when predicting the swap number, due to Sycamore's high connectivity and large physical qubit count, which bring Sycamore closer to the ideal scenario compared to other quantum computers. Moreover, across all quantum computers, circuit depth and circuit width consistently emerge as key features and demonstrating a complementary relationship. This underscores the importance of fundamental circuit characteristics in features. For IBM-Rochester and Aspen-4, circuit width and operation density are the most important features, with similar importance distributions, likely due to their lower connectivity. In these weaker-connected systems, increases in circuit width and operation density lead to larger circuit sizes and higher gate densities, which, in turn, require more swap operations. Overall, each feature shows strong importance on at least one quantum computer architecture, which is consistent with our considerations for feature selection.

\section{Machine Learning Quantum Dataset}
\label{app:mlqd}
This study introduces MLQD, an open-source dataset designed for qubit mapping research. Utilizing the OLSQ2 mapping methodology\cite{Lin2023}, MLQD comprehensively maps five quantum computer architectures: Sycamore, IBM Rochester, IBM Melbourne, Rigetti Aspen-4, and a 5x5 architecture.

The dataset comprises two circuit categories sourced from QASMbench\cite{Li2023} and our proposed enhancement method, including circuits with exclusively two-qubit gates and hybrid circuits integrating single- and two-qubit gates. Each architecture contains a minimum of 700 samples. Each sample consists of three files: pre-mapping and post-mapping quantum circuits (.qasm), and a mapping process information file (.json) detailing circuit depth and swap numbers. Post-mapped circuits are stored in the respective sample's /result folder.

MLQD facilitates research on quantum circuit mapping across diverse architectures and circuit characteristics. The dataset is continuously updating\cite{github_repo}.

\bibliographystyle{elsarticle-num}
\bibliography{mylib}

\end{document}